\documentstyle[12pt,epsfig]{article}
\begin{document}
\def\e{\epsilon}
\def\ycut{y_{\rm cut}}
\def\ymin{y_{\rm min}}
\def\yqqb{y_{q\bar{q}}}
\def\yqg{y_{q\gamma}}
\def\yqbg{y_{\bar{q}\gamma}}
\def\as{\left(\frac{\alpha_s}{2\pi}\right)}
\def\asN{\left(\frac{\alpha_sN}{2\pi}\right)}
\def\asmu{\left(\frac{\alpha_s(\mu)}{2\pi}\right)}
\def\asNmu{\left(\frac{\alpha_s(\mu)N}{2\pi}\right)}
\def\4pi2{\left(\frac{4\pi\mu^2}{M^2}\right)^{2 \e}}
\def\g2No2{\left(\frac{g^2N}{2}\right)}
\def\e{\epsilon}
\def\A{{\cal A}}
\def\B{{\cal B}}
\def\K{{\cal K}}
\def\F{{\cal F}}
\def\M{{\cal M}}
\def\S{{\cal S}}
\def\T{{\cal T}}
\def\I{{\cal I}}
\def\J{{\cal J}}
\def\L{{\cal L}}
\def\P{{\cal P}}
\def\R{{\cal R}}
\def\V{{\cal V}}
\def\Lhat{\hat {\cal L}}
\def\d{{\rm d}}
\def\Qbar{{\overline Q}}
\def\Li{{\rm Li}}


\begin{titlepage} 
\vspace*{-1cm} 
\begin{flushright} 
DTP/98/58   \\ 
September 1998 \\
\end{flushright}
\vskip 1.cm
\begin{center}                                                             
{\Large\bf Four jet event shapes in electron-positron annihilation} 
\vskip
1.3cm 
{\large J.~M.~Campbell, M.~A.~Cullen and E.~W.~N.~Glover} 
\vskip .2cm 
{\it Department of Physics, 
University of Durham, 
Durham DH1 3LE, 
England } 
\vskip
2.3cm    
\end{center}       
\begin{abstract} 

We report next-to-leading order perturbative QCD predictions of 4 jet event
shape variables for the process $e^+e^-\to\rm{4\,\,jets}$ obtained using the
general purpose Monte Carlo {\tt EERAD2}. This program is based on the known 
`squared' one loop matrix elements for the virtual $\gamma^* \to 4$ parton
contribution and squared matrix elements for 5 parton production. To combine
the two distinct final states numerically we present a hybrid of the commonly
used subtraction and slicing schemes based on the colour antenna structure of
the final state which can be readily applied to other processes. We have
checked that the numerical results obtained with {\tt EERAD2}  are consistent
with next-to-leading order estimates of the distributions of previously
determined four jet-like event variables. We also report the next-to-leading
order scale independent coefficients for some previously uncalculated
observables; the light hemisphere mass, narrow jet broadening, Aplanarity  and
the 4 jet transition variables with respect to the JADE and Geneva jet finding
algorithms. For each of these observables, the  next-to-leading order
corrections calculated at the physical scale significantly increase the rate
compared to leading order (the K factor is approximately 1.5 -- 2).  With the
exception of the 4 jet transition variables, the published DELPHI data lies
well above the  ${\cal O}(\alpha_s^3)$ predictions. The renormalisation scale
uncertainty is still large and in most cases the data prefers a scale
significantly smaller than the physical scale. This situation is reminiscent of
that for three jet shape variables, and should be improved by the inclusion of
power corrections and resummation of large infrared logarithms.

\end{abstract}                                                                
\vfill
\end{titlepage}
\newpage                                                                     
\section{Introduction} 
\setcounter{equation}{0}
\label{sec:intro}

Electron-positron colliders, in particular those at both LEP and SLAC,
have provided much precision data with which to probe the structure of
QCD.  This is particularly valuable data because the strong
interactions occur only in the final state and are not entangled with
the parton density functions associated with beams of hadrons. In
addition to measuring multi-jet production rates, more specific
information about the topology of the events can be extracted. To this
end, many variables have been introduced which characterize the
hadronic  structure of an event. For example, we can ask how planar or
how collimated an event is.  In general, a variable is described as $n$
jet-like if it vanishes for a final state configuration of $n-1$
hadrons. With the precision data from LEP and SLC, experimental
distributions for such event shape variables have been studied and have
been compared where possible with theoretical calculations. 

Generally speaking, leading order (LO) predictions successfully predict
the general features of distributions, but can be improved by resumming
kinematically-dominant logarithms, by including more perturbative
information or both. A next-to-leading order (NLO) treatment of
three-jet like variables was first performed in~\cite{ert,slice} and
systematically completed in~\cite{event}. Armed with such calculations,
one can extract a value for the strong coupling $\alpha_s$ either
directly from the event shape distributions~\cite{alphas1} or from the
energy dependence of their average value~\cite{alphas2}. Alternatively,
one can study the group parameters of the gauge theory of the strong
interactions \cite{3jetCF},  though for three jet observables, the
gluon-gluon coupling (proportional to $C_A$) occurs first at NLO.

Recent calculations of the relevant one-loop four parton matrix
elements for  $\gamma^* \to 4$~partons~\cite{us} and $e^+e^- \to
4$~partons \cite{them}, together with the known tree-level five parton
matrix elements~\cite{5part} have enabled the phenomenology of four-jet
production to be be studied at next-to-leading order.  So far, two
groups have used these matrix elements to construct general purpose 
Monte Carlo programs for four jet-like quantities {\tt
MENLO~PARC}~\cite{menlo-parc} and {\tt DEBRECEN}~\cite{debrecen}, which
have been used to measure the four jet fraction $R_4$ and a variety of
event shape distributions~\cite{debrecen}. We note that four jet
production is sensitive to the casimir structure of QCD 
\cite{NTgroup,them} and four jet events may be used to constrain the
allowed values of $C_F$, $C_A$ and $T_R$ \cite{NTcharge} by examining
the angles between the jets \cite{4jetangles}. One may also place
next-to-leading order bounds on the  possible presence of the elusive
light gluino present in many supersymmetric models \cite{NTgluino}.

Any such Monte Carlo program must suitably combine the virtual four parton and
real five parton pieces of the cross section. That is,  \begin{equation}
\sigma_{\rm NLO} = \int \d \sigma_4 + \int \d \sigma_5 = \int \d PS_4 \, |
{\cal M}_4 |^2 + \int \d PS_5 \, | {\cal M}_5 |^2,  \end{equation}  where the
$n$-parton contributions $\d \sigma_n$ are integrals over the $n$-parton
phase-space $\d PS_n$. Both integrals are separately divergent, and contain
both infrared and ultraviolet singularities.  The ultraviolet poles are removed
by renormalisation, however the soft and collinear infrared poles are only
cancelled when the virtual graphs are combined with the bremstrahlung process
and one must provide a means to cancel the infrared singularities caused by two
particles becoming collinear or one soft.   Although the cancellation of
infrared poles can be done analytically for simple processes, for complicated
processes, it is necessary to resort to numerical techniques.   A variety of
methods - known as subtraction~\cite{ert,subtract,event2},
slicing~\cite{slice,GG} and hybrid subtraction~\cite{hybrid} - are in general
use, and   variations of the subtraction formalism have been implemented
in~\cite{menlo-parc,debrecen}. Here we report on results obtained using a third
numerical implementation of these matrix elements to compute generic infrared
safe four jet observables. We call this program {\tt EERAD2} \cite{eerad2} and
it is based on the `squared' one-loop matrix elements for $\gamma^* \to
4$~partons of \cite{us} together with squared tree level matrix elements for
$\gamma^* \to 5$~partons. To isolate and  cancel the infrared singularities, we
use a hybrid approach that contains elements  of both slicing and subtraction
methods.   In particular, we use an antenna factorisation where two (colour
connected) hard partons radiate a third that may be unresolved. This is
necessarily a rather technical subject and it is detailed in the appendix.

For the bulk of the paper we are more concerned with the phenomenology of four
jet-like shape variables and, in particular,   we extend the set of variables
that have been computed at ${\cal O}(\alpha_s^3)$.\footnote{Some of the
results  presented here have been reported in \cite{moriond}.} To be precise, 
we present next-to-leading order coefficents for the differential distributions
of the narrow jet broadening, light hemisphere mass, Aplanarity and the jet
transition variable for the JADE and Geneva jet algorithms. To compare with
previous results, we also consider the thrust minor, $D$ parameter and the jet
transition variable for the Durham jet algorithm as well as the four jet rate
as a function of the jet resolution parameter $\ycut$. 

In section 2, we give the definitions  of the relevant four jet shape
variables and review the structure of the perturbative predictions.
Section 3 shows the consistency of our program, by reproducing the
known four-jet fraction and $D$ parameter distributions. The main
results are reported in section 4 and comparisons with experimental
data follow in section 5, where we also discuss how we might optimize
our perturbative input by choice of a suitable renormalization scale.
Conclusions are summarized in section 6. Finally, a more detailed
discussion of our method for cancellation of the infrared singularities
is reserved for the Appendix.

\newpage

\section{Four jet event shapes}
\setcounter{equation}{0}
\label{sec:shape}

The sorts of variables we are interested in are four jet-like, since
they can only be non-zero for final states in which there are four or
more particles. They usually rely on the hadronic final state having
some volume and, when the event is coplanar, some observables like the
$D$ parameter are identically zero.

\subsection{Definition of Variables}
\label{subsec:vardef}

In the following definitions, the sums run over all $N$ final state
particles, $k=1,\ldots,N$. $\vec{p}_k$ is the three-momentum of
particle $k$ in the c.m. frame, with  components $p_k^i$, $i=1,2,3$.

\begin{itemize}
\item[(a)] $C$ and $D$ parameters \cite{CDdef}. \vspace{0.25cm} \\
We first construct the linear momentum tensor,
\begin{equation}
\Theta^{ij} = \frac{\sum_k \frac{p^i_k p^j_k}{|\vec{p}_k|}}{\sum_k |\vec{p_k}|},
\end{equation}
with eigenvalues $\lambda_i$ for $i=1,2,3$. The normalisation is such
that $\sum_i \lambda_i = 1$. For planar events one of the eigenvalues
is zero. The $C$ and $D$ parameters are defined by,
\begin{equation}
D=27 \lambda_1\lambda_2\lambda_3,
\end{equation}
and,
\begin{equation}
C = 3(\lambda_1\lambda_2+\lambda_2\lambda_3 + \lambda_3\lambda_1).
\end{equation}
$D$ can only be non-zero for non-planar four (or more) parton events, 
while three parton events may produce $0 \leq C \leq 0.75$.  
Only the region $C > 0.75$ should be considered four jet-like. 
\item[(b)] Thrust minor, $T_{{\rm minor}}$ \cite{Tdef}. \vspace{0.25cm} \\
We first define the thrust, major and minor axes
($\vec{n}_1,\vec{n}_2,\vec{n}_3$) by,
\begin{equation}
T_i=\max_{\vec{n}_i} \frac{\sum_k |\vec{p}_k \cdot \vec{n}_k|}
 {\sum_k |\vec{p}_k|},
\end{equation}
where $\vec{n}_2$ is constrained by $\vec{n}_1 \cdot \vec{n}_2=0$.
and $\vec{n}_3 = \vec{n}_1 \times \vec{n}_2$. 
\item[(c)] Light hemisphere mass, $M_L^2/s$. \vspace{0.25cm} \\
The event is separated into two hemispheres $H_1$, $H_2$ divided by the plane
normal to the thrust axis $\vec{n}_1$, as defined above.  Particles that
satisfy $\vec{p}_i.\vec{n}_1 > 0$ are assigned to hemisphere $H_1$, while all
other particles are in $H_2$. Then,
\begin{equation}
\frac{M_L^2}{s} = \frac{1}{s} \cdot \min_{i=1,2}
 \left( \sum_{\vec{p}_k \in H_i} p_k \right)^2.
\end{equation}
Note that this is the common modification of the original definition suggested
by Clavelli \cite{clavelli}.
\item[(d)] Narrow jet broadening, $B_{{\rm min}}$ \cite{Bdef}. 
\vspace{0.25cm} \\
Using the same division into hemispheres as above, we define,
\begin{equation}
B_{\rm min}=\min_{i=1,2} \frac{\sum_{\vec{p}_k \in H_i} 
 |\vec{p}_k \times \vec{n}|}{2\sum_k|\vec{p}_k|}.
\end{equation}
\item[(e)] Aplanarity, $A$ \cite{Adef}. \vspace{0.25cm} \\
Here we consider the eigenvalues,  $\lambda_1 \geq \lambda_2 \geq
\lambda_3$ of 
the quadratic momentum tensor,
\begin{equation}
\Phi^{ij} = \frac{\sum_k p^i_k p^j_k}{\sum_k |\vec{p}_k|^2}.
\end{equation}
Aplanarity is defined by,
\begin{equation}
A=\frac{3}{2}\lambda_3,
\end{equation}
which is clearly zero for planar three particle events.
\item[(f)] Jet transition variable $y_4^S$. \vspace{0.25cm} \\ 
The $y_4^S$ variable denotes the value of the jet resolution parameter  $\ycut$
at which an event changes from a four jet event to a three jet event where the
jets are defined according to algorithm $S$. We consider three algorithms, the
JADE algorithm ($S=J$) \cite{jade}, the Durham  algorithm ($S=D$) \cite{durham} and the
Geneva algorithm ($S=G$)  \cite{geneva}. The jet-finding measures for each of
these three algorithms are as follows,
\begin{eqnarray}
& &y^J_{ij}=\frac{2\,E_iE_j(1-\cos\theta_{ij})}{s}\nonumber,\\
& &y^D_{ij}=\frac{2\,\min(E_i^2,E_j^2)(1-\cos\theta_{ij})}{s}\nonumber,\\
& &y^G_{ij}=\frac{8}{9}\frac{E_iE_j(1-\cos\theta_{ij})}{(E_i+E_j)^2},
\label{eq:jetmeasure}
\end{eqnarray} 
where the factor of $8/9$ in the Geneva algorithm is simply to ensure that the
maximum value of $y_{\rm{cut}}$ that reconstructs three jets from three partons
is $1/3$ as it is for the other two algorithms. When particles combine, there
is some ambiguity as to how to add the energies and momenta. In all three
schemes, we use the E scheme i.e. we
merely add four momenta,
\begin{equation}
p_{ij}^\mu = p_i^\mu + p_j^\mu.
\end{equation}
Other choices such as the E0 or P schemes 
where the cluster is made massless by 
rescaling the momentum or energy give similar results.  

\end{itemize}

Of these variables, the $D$, $C$, $T_{\rm minor}$ and $y_4^D$ distributions have been studied 
in \cite{debrecen}.

\subsection{Structure of Perturbative Prediction}
\label{subsec:structure}

The differential cross-section at centre-of-mass energy $\sqrt{s}$ for one of
these four-jet variables ($O_4$) at next-to-leading order is described by two
coefficients, $B_{O_4}$ and $C_{O_4}$ which represent the leading and
next-to-leading order perturbative contributions,
\begin{equation}
\frac{1}{\sigma_0} \cdot O_4 \, \frac{d\sigma}{dO_4} =
 \left( \frac{\alpha_s(\mu)}{2\pi} \right)^2 B_{O_4}
+\left( \frac{\alpha_s(\mu)}{2\pi} \right)^3 \left( 2 \beta_0
 \log \left(\frac{\mu^2}{s}\right) B_{O_4} + C_{O_4} \right).
\label{eq:def}
\end{equation}
Both $B_{O_4}$ and $C_{O_4}$ are scale independent and do not depend on the
beam energy. However, the running coupling $\alpha_s$ is calculated at
renormalization scale $\mu$ which is commonly chosen to be the  {\em physical}
scale, $\mu = \sqrt{s}$. Compared to the leading order prediction, which decays
monotonically with increasing $\mu$, the next-to-leading order term reduces the
scale dependence somewhat  through the first coefficient of the beta-function,
$\beta_0 = (33 - 2N_f) / \, 6$. For five active quark flavours, $\beta_0 =
3.833$.

The individual coefficients $B_{O_4}$ and $C_{O_4}$ depend on the number of
colours and flavours, or equivalently, the group casimirs of the standard
model. As such, four jet event shapes may be used to simultaneously  constrain
the strong interaction gauge group as well as the strong
coupling constant.

\subsection{Scale choice and theoretical uncertainty}
\label{subsec:scales}

As mentioned above, for hadronic observables in electron-positron annihilation
it is common to choose the renormalisation scale to be the physical scale $\mu
= \sqrt{s}$.  This choice is motivated by naturalness arguments and the fact
that choosing a scale far from $\sqrt{s}$ introduces large logarithms of  the
form $\log(\mu/\sqrt{s})$ in eq.~(\ref{eq:def}).   Since the renormalisation
scale is only a theoretical construct, 
it has become common practice to estimate the
uncertainty engendered by truncating the perturbative expansion by varying
$\mu$ by factors about the physical scale.

Other approaches have been considered with rather different scale choices.  One
such approach is to stipulate that the next-to-leading order coefficient
vanishes.  That is to say that the LO and NLO predictions 
coincide.\footnote{In other words, the ratio of NLO to LO predictions commonly
called the K-factor is identically unity.} This occurs at the scale,
\begin{equation}
\mu^{FAC} = \sqrt{s} ~ \exp\left(-\frac{C_{O_4}}{4B_{O_4}\beta_0}\right),
\label{eq:facscale}
\end{equation}
and is called the scale of {\em fastest apparent convergence} (FAC) \cite{FAC}.
As we will see, $C_{O_4}$ is typically 50-100 times larger than $B_{O_4}$ so 
that $\mu^{FAC}$ may be as small as $10^{-2}$--$10^{-3}$ of the physical scale.

Another common choice is to select the scale where the NLO prediction is most
insensitive to the choice of scale \cite{PMS}.   However, this scale is very
close to the FAC scale (15\% smaller) so that the {\em Principle of Minimal
Sensitivity} scale (PMS) predictions lie very close to the FAC scale
prediction.

While even higher order corrections remain uncalculated, varying the
renormalisation scale can only give a crude indication of the theoretical
uncertainty. Therefore, in an attempt to make a fair estimate of the
theoretical uncertainty on the NLO prediction we will show both the physical
scale and FAC scale predictions.

\subsection{Infrared behaviour} \label{subsec:IRbehaviour}

Four jet event shapes typically depend on the event having some volume and not
lying entirely in a plane. Typical hadronic events contain more than 20 hadrons
and it is extremely unlikely that the value of any event shape is precisely
zero for any experimental event.However, in a LO or NLO fixed order parton
calculation,  there only four or five partons present in the final state and,
when one or more are soft, the calculated $O_4$ may approach zero. In such
circumstances, soft gluon singularities cause the fixed order prediction to
become wildly unstable and grow logarithmically. In the small $O_4$ limits, the
perturbative coefficients have the following form,
\begin{eqnarray}
B_{O_4} &\to&  A_{32} L^3+A_{22} L^2+A_{12} L+A_{02},\nonumber \\
C_{O_4} &\to&  A_{53} L^5+A_{43} L^4+A_{33} L^3+A_{23}L^2+A_{13} L+A_{03},
\label{eq:logbehaviour}
\end{eqnarray}
where $L = \log(1/O_4)$ and $A_{nm}$ are (as yet) undetermined coefficients.
Whenever $L$ is sufficiently large, resummation effects will be
important.\footnote{Whether the coefficients exponentiate and can be resummed
will depend on the observable.} In comparing with data, we therefore choose to
make a cut on the size of  $O_4$ which is typically in the range 0.001 ---
0.01, since for such small values of $O_4$ we do not trust the NLO
prediction.   In comparing with the DELPHI data, this cut will usually be the
lower edge of the second data bin.

\newpage

\section{Comparison with existing results}
\setcounter{equation}{0}
\label{sec:compare}

\subsection{Four jet rates}

As a check of the numerical results, Table~\ref{table} shows the
predictions for each of the three Monte Carlo programs for the four jet rate
for three jet clustering algorithms;
the Jade-E0,  Durham-E \cite{durham},  and Geneva-E \cite{geneva}
algorithms.
We show results with $\alpha_s(M_Z)=0.118$ 
for three values of the jet resolution parameter $y_{cut}$.
There is good agreement with the results from the other two calculations.

\begin{table}[h]
\begin{center}
\begin{tabular}{|c|c|c|c|c|}
\hline
Algorithm & $y_{{\rm cut}}$ & {\tt MENLO PARC} & {\tt DEBRECEN} & {\tt EERAD2} \\ \hline
\hline
         & 0.005& $ (1.04\pm0.02) \cdot 10^{-1} $ & $ (1.05\pm0.01) \cdot
10^{-1} $ & $ (1.05\pm0.01) \cdot 10^{-1} $\\
Durham & 0.01 & $ (4.70\pm0.06) \cdot 10^{-2} $ & $ (4.66\pm0.02) \cdot
10^{-2} $ & $ (4.65\pm0.02) \cdot 10^{-2} $\\
       & 0.03 & $ (6.82\pm0.08) \cdot 10^{-3} $ & $ (6.87\pm0.04) \cdot
10^{-3} $ & $ (6.86\pm0.03) \cdot 10^{-3} $\\
\hline
\hline
       & 0.02 & $ (2.56\pm0.06) \cdot 10^{-1} $ & $ (2.63\pm0.06) \cdot
10^{-1} $ & $ (2.61\pm0.05) \cdot 10^{-1} $\\
Geneva & 0.03 & $ (1.71\pm0.03) \cdot 10^{-1} $ & $ (1.75\pm0.03) \cdot
10^{-1} $ & $ (1.72\pm0.03) \cdot 10^{-1} $\\
       & 0.05 & $ (8.58\pm0.15) \cdot 10^{-2} $ & $ (8.37\pm0.12) \cdot
10^{-2} $ & $ (8.50\pm0.06) \cdot 10^{-2} $\\
\hline
\hline
       & 0.005& $ (3.79\pm0.08) \cdot 10^{-1} $ & $ (3.88\pm0.07) \cdot
10^{-1} $ & $ (3.87\pm0.03) \cdot 10^{-1} $\\
JADE-E0 & 0.01& $ (1.88\pm0.03) \cdot 10^{-1} $ & $ (1.92\pm0.01) \cdot
10^{-1} $ & $ (1.93\pm0.01) \cdot 10^{-1} $\\
       & 0.03 & $ (3.46\pm0.05) \cdot 10^{-2} $ & $ (3.37\pm0.01) \cdot
10^{-2} $ & $ (3.35\pm0.01) \cdot 10^{-2} $\\
\hline
\end{tabular}
\caption{The four-jet fraction as calculated by {\tt MENLO~PARC},
{\tt DEBRECEN} and {\tt EERAD2},
for the different jet recombination schemes and varying $y_{\rm cut}$.
The rate is normalized by the ${\cal O}(\alpha_s)$
total hadronic cross-section, $\sigma_{\rm had}=\sigma_0 \, (1+
\alpha_s/\pi)$.}
\label{table}
\end{center}
\end{table}

\subsection{Shape variables}

As mentioned earlier, Nagy and Tr{\'o}cs{\'a}nyi \cite{debrecen} have computed
$C_D$ with their Monte Carlo {\tt DEBRECEN}. In Table~\ref{tab:Dpar} we show
the leading and next-to-leading order coefficients $B_D$ and $C_D$ calculated
by {\tt EERAD2}, together with the {\tt DEBRECEN} result. The two calculations
are clearly consistent with one another, with the quoted errors overlapping in
almost all cases. The errors from {\tt EERAD2} are of the order of 2\% in each
bin, except in the tail of the distribution where the errors rise as high as
10\%. The infrared enhancement of the distribution described in
section~\ref{subsec:IRbehaviour} means that the Monte Carlo procedure favours
the phase space region corresponding to small values of the $D$ parameter, so
that the large $D$ tail suffers larger errors. In fact $C_D$ drops by four
orders of magnitude over the kinematic range of the observable so it is
necessary to use importance sampling with respect to  the observable
distribution to ensure sufficient Monte Carlo points  are produced in the high
$D$ region.  This is also true for all of the other shape variables.

In addition, Nagy and Tr{\'o}cs{\'a}nyi have also presented results for the 
next-to-leading order coefficents for thrust minor $T_{\rm minor}$ and the
jet transition variable in the Durham scheme $y_4^D$~\cite{debrecen}. Although we
do not present a detailed comparison here, we note that the agreement is
qualitatively the same as discussed for the $D$ parameter above. We find that
the distributions extend beyond the range of coefficents presented
in~\cite{debrecen}, with non-zero coefficients for bins in the ranges $0.5 < T_{\rm
minor} < 0.58$ and $0.125 < y_4^D < 0.17$.

\begin{table}[h]
\begin{center}
\begin{tabular}{|c|c|c|c|}
\hline
$D$ & $B_{D}$ & $C_{D}$ & {\tt DEBRECEN}\\ 
\hline \hline
0.0200&$( 3.79 \pm  0.01) \cdot 10^{ 2}$&$( 1.47 \pm  0.00) \cdot 10^{ 4}$&$( 1.08 \pm  0.06) \cdot 10^{ 4}$  \\
0.0600&$( 2.32 \pm  0.01) \cdot 10^{ 2}$&$( 1.25 \pm  0.01) \cdot 10^{ 4}$&$( 1.24 \pm  0.02) \cdot 10^{ 4}$  \\
0.1000&$( 1.45 \pm  0.01) \cdot 10^{ 2}$&$( 8.69 \pm  0.04) \cdot 10^{ 3}$&$( 8.59 \pm  0.12) \cdot 10^{ 3}$  \\
0.1400&$( 1.04 \pm  0.01) \cdot 10^{ 2}$&$( 6.39 \pm  0.03) \cdot 10^{ 3}$&$( 6.24 \pm  0.12) \cdot 10^{ 3}$  \\
0.1800&$( 7.68 \pm  0.04) \cdot 10^{ 1}$&$( 4.89 \pm  0.03) \cdot 10^{ 3}$&$( 4.99 \pm  0.11) \cdot 10^{ 3}$  \\
0.2200&$( 5.87 \pm  0.03) \cdot 10^{ 1}$&$( 3.88 \pm  0.03) \cdot 10^{ 3}$&$( 3.85 \pm  0.06) \cdot 10^{ 3}$  \\
0.2600&$( 4.66 \pm  0.07) \cdot 10^{ 1}$&$( 3.04 \pm  0.03) \cdot 10^{ 3}$&$( 2.98 \pm  0.05) \cdot 10^{ 3}$  \\
0.3000&$( 3.75 \pm  0.07) \cdot 10^{ 1}$&$( 2.51 \pm  0.04) \cdot 10^{ 3}$&$( 2.52 \pm  0.05) \cdot 10^{ 3}$  \\
0.3400&$( 3.07 \pm  0.05) \cdot 10^{ 1}$&$( 2.02 \pm  0.03) \cdot 10^{ 3}$&$( 1.94 \pm  0.05) \cdot 10^{ 3}$  \\
0.3800&$( 2.41 \pm  0.03) \cdot 10^{ 1}$&$( 1.61 \pm  0.03) \cdot 10^{ 3}$&$( 1.59 \pm  0.04) \cdot 10^{ 3}$  \\
0.4200&$( 1.97 \pm  0.04) \cdot 10^{ 1}$&$( 1.37 \pm  0.02) \cdot 10^{ 3}$&$( 1.37 \pm  0.03) \cdot 10^{ 3}$  \\
0.4600&$( 1.56 \pm  0.03) \cdot 10^{ 1}$&$( 1.09 \pm  0.01) \cdot 10^{ 3}$&$( 1.06 \pm  0.03) \cdot 10^{ 3}$  \\
0.5000&$( 1.32 \pm  0.01) \cdot 10^{ 1}$&$( 8.97 \pm  0.14) \cdot 10^{ 2}$&$( 8.72 \pm  0.19) \cdot 10^{ 2}$  \\
0.5400&$( 1.05 \pm  0.02) \cdot 10^{ 1}$&$( 7.12 \pm  0.15) \cdot 10^{ 2}$&$( 7.11 \pm  0.16) \cdot 10^{ 2}$  \\
0.5800&$( 8.46 \pm  0.16) \cdot 10^{ 0}$&$( 5.79 \pm  0.12) \cdot 10^{ 2}$&$( 5.68 \pm  0.14) \cdot 10^{ 2}$  \\
0.6200&$( 6.60 \pm  0.16) \cdot 10^{ 0}$&$( 4.55 \pm  0.09) \cdot 10^{ 2}$&$( 4.46 \pm  0.21) \cdot 10^{ 2}$  \\
0.6600&$( 5.32 \pm  0.13) \cdot 10^{ 0}$&$( 3.58 \pm  0.07) \cdot 10^{ 2}$&$( 3.52 \pm  0.11) \cdot 10^{ 2}$  \\
0.7000&$( 3.99 \pm  0.09) \cdot 10^{ 0}$&$( 2.80 \pm  0.09) \cdot 10^{ 2}$&$( 2.74 \pm  0.09) \cdot 10^{ 2}$  \\
0.7400&$( 3.06 \pm  0.05) \cdot 10^{ 0}$&$( 2.05 \pm  0.08) \cdot 10^{ 2}$&$( 2.08 \pm  0.08) \cdot 10^{ 2}$  \\
0.7800&$( 2.26 \pm  0.04) \cdot 10^{ 0}$&$( 1.58 \pm  0.04) \cdot 10^{ 2}$&$( 1.54 \pm  0.06) \cdot 10^{ 2}$  \\
0.8200&$( 1.54 \pm  0.04) \cdot 10^{ 0}$&$( 1.05 \pm  0.03) \cdot 10^{ 2}$&$( 1.03 \pm  0.04) \cdot 10^{ 2}$  \\
0.8600&$( 9.72 \pm  0.21) \cdot 10^{-1}$&$( 6.72 \pm  0.29) \cdot 10^{ 1}$&$( 6.66 \pm  0.31) \cdot 10^{ 1}$  \\
0.9000&$( 5.63 \pm  0.16) \cdot 10^{-1}$&$( 3.85 \pm  0.17) \cdot 10^{ 1}$&$( 3.89 \pm  0.20) \cdot 10^{ 1}$  \\
0.9400&$( 2.62 \pm  0.07) \cdot 10^{-1}$&$( 1.71 \pm  0.10) \cdot 10^{ 1}$&$( 1.71 \pm  0.19) \cdot 10^{ 1}$  \\
0.9800&$( 5.34 \pm  0.11) \cdot 10^{-2}$&$( 3.15 \pm  0.27) \cdot 10^{ 0}$&$( 2.60 \pm  1.30) \cdot 10^{ 0}$  \\
\hline
\end{tabular}
\caption{The leading and next-to-leading order coefficients for the $D$
parameter.
The NLO coefficient predicted by Nagy and Tr{\'o}cs{\'a}nyi Monte Carlo {\tt DEBRECEN}
\cite{debrecen} is also shown.}
\label{tab:Dpar}
\end{center}
\end{table}

\newpage

\section{New results}
\setcounter{equation}{0}
\label{sec:results}

In this section we extend the analysis of 4 jet-like event shape observables
already found in the literature by reporting the leading and next-to-leading
order coefficients for the light hemisphere mass, the narrow hemisphere
broadening, Aplanarity and the jet transition variable in both the JADE and
Geneva schemes, $y_4^J$  and $y_4^G$. In particular, we examine the relative
sizes of the two terms by inspecting the K factor (at the physical scale) for
each variable across the allowed kinematic range of the distributions.

For all the variables presented in this section, we must be careful to
differentiate between the true behaviour of the distribution as the observable
tends to zero and the behaviour in fixed order perturbation theory. Each of the
observables should have a smooth behaviour as $O_4 \to 0$ rather than the
divergent behaviour exhibited by the coefficients according to
equation~\ref{eq:logbehaviour}. To recover a smooth result in this limit it is
necessary to resum powers of $\log(1/O_{4}$) where possible, a procedure which
has been performed already for many 3 jet-like variables \cite{catani,Bdef}

\subsection{Light Hemisphere Mass}

\begin{table}[h]
\begin{center}
\begin{tabular}{|c|c|c|}
\hline
$M_L^2/s$ & $B_{M_L^2/s}$ & $C_{M_L^2/s}$ \\ 
\hline \hline
0.0150&$( 3.23 \pm  0.08) \cdot 10^{ 2}$&$( 1.41 \pm  0.01) \cdot 10^{ 4}$  \\
0.0250&$( 1.88 \pm  0.02) \cdot 10^{ 2}$&$( 8.85 \pm  0.10) \cdot 10^{ 3}$  \\
0.0350&$( 1.25 \pm  0.02) \cdot 10^{ 2}$&$( 5.97 \pm  0.11) \cdot 10^{ 3}$  \\
0.0450&$( 8.52 \pm  0.10) \cdot 10^{ 1}$&$( 4.14 \pm  0.08) \cdot 10^{ 3}$  \\
0.0550&$( 5.97 \pm  0.06) \cdot 10^{ 1}$&$( 3.04 \pm  0.04) \cdot 10^{ 3}$  \\
0.0650&$( 4.20 \pm  0.09) \cdot 10^{ 1}$&$( 2.15 \pm  0.05) \cdot 10^{ 3}$  \\
0.0750&$( 3.02 \pm  0.07) \cdot 10^{ 1}$&$( 1.58 \pm  0.05) \cdot 10^{ 3}$  \\
0.0850&$( 2.13 \pm  0.03) \cdot 10^{ 1}$&$( 1.11 \pm  0.02) \cdot 10^{ 3}$  \\
0.0950&$( 1.39 \pm  0.04) \cdot 10^{ 1}$&$( 7.66 \pm  0.23) \cdot 10^{ 2}$  \\
0.1050&$( 8.75 \pm  0.20) \cdot 10^{ 0}$&$( 4.97 \pm  0.17) \cdot 10^{ 2}$  \\
0.1150&$( 5.18 \pm  0.13) \cdot 10^{ 0}$&$( 3.27 \pm  0.07) \cdot 10^{ 2}$  \\
0.1250&$( 2.59 \pm  0.12) \cdot 10^{ 0}$&$( 1.66 \pm  0.07) \cdot 10^{ 2}$  \\
0.1350&$( 8.97 \pm  0.35) \cdot 10^{-1}$&$( 6.61 \pm  0.41) \cdot 10^{ 1}$  \\
0.1450&$( 2.49 \pm  0.13) \cdot 10^{-1}$&$( 1.79 \pm  0.09) \cdot 10^{ 1}$  \\
0.1550&$( 5.00 \pm  0.27) \cdot 10^{-2}$&$( 3.75 \pm  0.26) \cdot 10^{ 0}$  \\
0.1650&$( 1.46 \pm  0.21) \cdot 10^{-3}$&$( 2.30 \pm  0.37) \cdot 10^{-1}$  \\
\hline
\end{tabular}
\caption{The leading and next-to-leading order coefficients for the light jet
mass $M_L^2/s$.}
\label{table:Light}
\end{center}
\end{table}

\begin{figure}[h]
\begin{center}
\psfig{figure=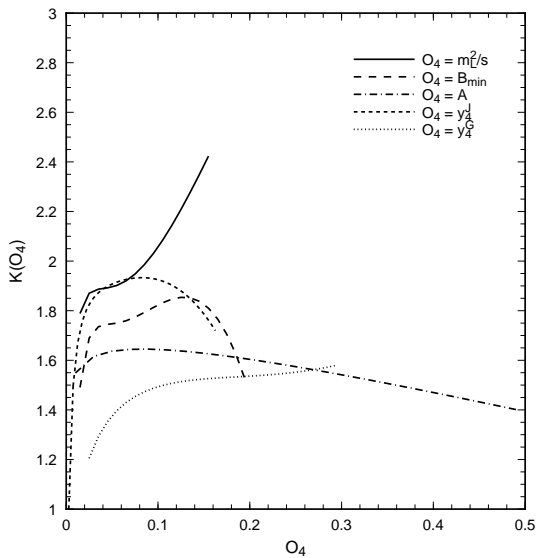,width=8cm}
\end{center}
\caption[]{The K-factors defined according to eq.~(\ref{eq:Kfac})
for four jet event shapes,
the light hemisphere mass (solid), narrow jet broadening (long-dashed),
Aplanarity (dot-dashed)
and jet transition variables in the JADE (short-dashed) and Geneva (dotted)
schemes.  Each variable has a different kinematic range.}
\label{fig:Kfac}
\end{figure}

As defined before, the light hemisphere mass is the smaller invariant mass of
the two hemispheres formed by separating the event by a plane normal to the
thrust axis.  The NLO coefficient $C_{M_L^2/s}$ evaluted at the physical
scale $\mu = \sqrt{s}$  together with the LO term is given in
Table~\ref{table:Light}.   The errors are estimates from the numerical program
and are typically 2-3\% for each entry. As with the previously known results on
four jet event shapes, the NLO terms are significantly larger than the LO
term.   Here, we see that $C_{M_L^2/s}$  is typically 50 times larger than
$B_{M_L^2/s}$ so that even when the additional factor of $\alpha_s/2\pi$ is
restored, the NLO correction is large. This is illustrated in
Fig.~\ref{fig:Kfac} where the K factor defined by,
\begin{equation}
K_{O_4} = 
1+\left(\frac{\alpha_s(\sqrt{s})}{2\pi} \right)\frac{ C_{O_4}}{B_{O_4}},
\label{eq:Kfac}
\end{equation}
is shown for $O_4 = M_L^2/s$. We see that the K factor increases with the value
of the observable, rising from 1.8 at small $M_L^2/s$ up to 2.4. This behaviour
is similar to that observed for other four jet event shapes
\cite{debrecen}.

\subsection{Narrow Hemisphere Broadening}

Narrow hemisphere broadening, $B_{\rm min}$, is defined in a  similiar manner
to the light hemisphere mass. The event is again divided into two hemispheres
by the plane normal to the thrust axis, but now the momenta transverse to the
thrust axis is summed (normalised by the sum of absolute momenta) in each
hemisphere. The narrow hemisphere is that with the least transverse momentum
with respect to the thrust axis. Numerical results for this variable as
calculated by {\tt EERAD2}  can be found in Table~\ref{table:Bmin}. As with the
light hemisphere mass, the NLO contribution is significant yielding a K factor
of roughly 1.7 over most of the kinematic range of the variable (see
Fig.~\ref{fig:Kfac}).

\begin{table}[h] 
\begin{center}
\begin{tabular}{|c|c|c|} 
\hline $B_{\min}$ & $B_{B_{\min}}$ & $C_{B_{\min}}$
\\  \hline \hline 
0.0150&$( 1.19 \pm  0.01) \cdot 10^{ 3}$&$( 3.41 \pm  0.07) \cdot 10^{ 4}$  \\
0.0250&$( 7.04 \pm  0.06) \cdot 10^{ 2}$&$( 2.56 \pm  0.02) \cdot 10^{ 4}$  \\
0.0350&$( 4.80 \pm  0.02) \cdot 10^{ 2}$&$( 1.92 \pm  0.04) \cdot 10^{ 4}$  \\
0.0450&$( 3.39 \pm  0.02) \cdot 10^{ 2}$&$( 1.41 \pm  0.02) \cdot 10^{ 4}$  \\
0.0550&$( 2.49 \pm  0.02) \cdot 10^{ 2}$&$( 1.07 \pm  0.02) \cdot 10^{ 4}$  \\
0.0650&$( 1.89 \pm  0.02) \cdot 10^{ 2}$&$( 8.04 \pm  0.12) \cdot 10^{ 3}$  \\
0.0750&$( 1.43 \pm  0.02) \cdot 10^{ 2}$&$( 6.29 \pm  0.12) \cdot 10^{ 3}$  \\
0.0850&$( 1.08 \pm  0.01) \cdot 10^{ 2}$&$( 4.81 \pm  0.08) \cdot 10^{ 3}$  \\
0.0950&$( 8.19 \pm  0.04) \cdot 10^{ 1}$&$( 3.65 \pm  0.08) \cdot 10^{ 3}$  \\
0.1050&$( 6.23 \pm  0.08) \cdot 10^{ 1}$&$( 2.77 \pm  0.09) \cdot 10^{ 3}$  \\
0.1150&$( 4.69 \pm  0.06) \cdot 10^{ 1}$&$( 2.10 \pm  0.04) \cdot 10^{ 3}$  \\
0.1250&$( 3.37 \pm  0.04) \cdot 10^{ 1}$&$( 1.45 \pm  0.04) \cdot 10^{ 3}$  \\
0.1350&$( 2.36 \pm  0.04) \cdot 10^{ 1}$&$( 1.09 \pm  0.03) \cdot 10^{ 3}$  \\
0.1450&$( 1.64 \pm  0.03) \cdot 10^{ 1}$&$( 7.07 \pm  0.25) \cdot 10^{ 2}$  \\
0.1550&$( 9.82 \pm  0.12) \cdot 10^{ 0}$&$( 4.48 \pm  0.15) \cdot 10^{ 2}$  \\
0.1650&$( 5.08 \pm  0.12) \cdot 10^{ 0}$&$( 2.18 \pm  0.10) \cdot 10^{ 2}$  \\
0.1750&$( 1.71 \pm  0.04) \cdot 10^{ 0}$&$( 7.53 \pm  0.33) \cdot 10^{ 1}$  \\
0.1850&$( 4.32 \pm  0.11) \cdot 10^{-1}$&$( 1.59 \pm  0.12) \cdot 10^{ 1}$  \\
0.1950&$( 5.47 \pm  0.11) \cdot 10^{-2}$&$( 1.34 \pm  0.24) \cdot 10^{ 0}$  \\
\hline 
\end{tabular} 
\caption{The leading
and next-to-leading order coefficients for the narrow jet broadening
$B_{\min}$.} 
\label{table:Bmin} 
\end{center} 
\end{table}

\subsection{Aplanarity}

As described earlier, Aplanarity is essentially the smallest eigenvalue of the 
quadratic momentum tensor.
The NLO coefficient $C_{A}$ evaluted at the physical
scale $\mu = \sqrt{s}$  together with the LO term is given in
Table~\ref{table:Apar}.  Once again, the NLO terms 
are significantly larger than the LO
term $B_{A}$ and, as can be seen in Fig.~\ref{fig:Kfac} gives rise to K factor 
of roughly 1.5.

\begin{table}[h]
\begin{center}
\begin{tabular}{|c|c|c|}
\hline
$A$ & $B_{A}$ & $C_{A}$ \\ 
\hline \hline
0.0300&$( 7.25 \pm  0.04) \cdot 10^{ 1}$&$( 2.48 \pm  0.10) \cdot 10^{ 3}$  \\
0.0500&$( 3.87 \pm  0.03) \cdot 10^{ 1}$&$( 1.40 \pm  0.07) \cdot 10^{ 3}$  \\
0.0700&$( 2.35 \pm  0.02) \cdot 10^{ 1}$&$( 7.65 \pm  0.58) \cdot 10^{ 2}$  \\
0.0900&$( 1.57 \pm  0.02) \cdot 10^{ 1}$&$( 5.55 \pm  0.40) \cdot 10^{ 2}$  \\
0.1100&$( 1.06 \pm  0.01) \cdot 10^{ 1}$&$( 4.07 \pm  0.36) \cdot 10^{ 2}$  \\
0.1300&$( 7.45 \pm  0.12) \cdot 10^{ 0}$&$( 2.34 \pm  0.34) \cdot 10^{ 2}$  \\
0.1500&$( 5.19 \pm  0.07) \cdot 10^{ 0}$&$( 1.75 \pm  0.20) \cdot 10^{ 2}$  \\
0.1700&$( 3.76 \pm  0.08) \cdot 10^{ 0}$&$( 1.23 \pm  0.28) \cdot 10^{ 2}$  \\
0.1900&$( 2.68 \pm  0.04) \cdot 10^{ 0}$&$( 9.42 \pm  1.25) \cdot 10^{ 1}$  \\
0.2100&$( 2.02 \pm  0.03) \cdot 10^{ 0}$&$( 6.37 \pm  0.98) \cdot 10^{ 1}$  \\
0.2300&$( 1.32 \pm  0.02) \cdot 10^{ 0}$&$( 4.19 \pm  0.99) \cdot 10^{ 1}$  \\
0.2500&$( 9.70 \pm  0.16) \cdot 10^{-1}$&$( 3.14 \pm  0.67) \cdot 10^{ 1}$  \\
0.2700&$( 6.43 \pm  0.16) \cdot 10^{-1}$&$( 2.08 \pm  0.45) \cdot 10^{ 1}$  \\
0.2900&$( 4.28 \pm  0.14) \cdot 10^{-1}$&$( 1.25 \pm  0.34) \cdot 10^{ 1}$  \\
0.3100&$( 2.87 \pm  0.12) \cdot 10^{-1}$&$( 7.58 \pm  1.36) \cdot 10^{ 0}$  \\
0.3300&$( 1.79 \pm  0.03) \cdot 10^{-1}$&$( 6.13 \pm  1.67) \cdot 10^{ 0}$  \\
0.3500&$( 1.07 \pm  0.02) \cdot 10^{-1}$&$( 3.94 \pm  0.91) \cdot 10^{ 0}$  \\
0.3700&$( 6.66 \pm  0.20) \cdot 10^{-2}$&$( 2.00 \pm  0.40) \cdot 10^{ 0}$  \\
0.3900&$( 3.30 \pm  0.20) \cdot 10^{-2}$&$( 9.26 \pm  3.71) \cdot 10^{-1}$  \\
0.4100&$( 1.56 \pm  0.09) \cdot 10^{-2}$&$( 4.00 \pm  1.52) \cdot 10^{-1}$  \\
0.4300&$( 5.58 \pm  0.22) \cdot 10^{-3}$&$( 1.50 \pm  0.92) \cdot 10^{-1}$  \\
0.4500&$( 1.53 \pm  0.18) \cdot 10^{-3}$&$( 7.44 \pm  3.27) \cdot 10^{-2}$  \\
0.4700&$( 1.80 \pm  0.23) \cdot 10^{-4}$&$( 1.76 \pm  7.87) \cdot 10^{-3}$  \\
0.4900&$( 1.38 \pm  0.45) \cdot 10^{-5}$&$( 9.93 \pm 54.90) \cdot 10^{-5}$  \\
\hline
\end{tabular}
\caption{The leading and next-to-leading order coefficients for Aplanarity.}
\label{table:Apar}
\end{center}
\end{table}

\subsection{Jet transition variables}

As previously stated the jet transition variable $y_{4}^{S}$ describes the
scale where two jets merge, thereby changing a four jet event into a three jet
event. This is essentially the same as the derivative of the four jet rate with
respect to the jet resolution parameter $y_{\rm cut}$. However, the number of
jets in an event is dependent on the jet finding algorithm used to define the
\lq closeness' of particles which is compared with $y_{\rm cut}$. In \cite{debrecen}
the transition rate for the the Durham jet finding algorithm \cite{durham} is
given and we have checked that our results are consistent with these
predictions.  Here, we provide results for two other jet algorithms, the JADE
and Geneva \cite{geneva} schemes for which the jet finding measures are given
in eq.~(\ref{eq:jetmeasure}). We note that the Geneva algorithm enjoys the same
benefits as the Durham algorithm in that it is also supposed to exponentiate,
enabling infrared logarithms to be safely resummed. It also ensures that softly
radiated gluons are clustered with hard partons unless the angle of separation
between two soft gluons is much smaller  than the angular separation between
them and a hard parton.

Our results for the two schemes are given in Tables~\ref{table:y34G} and
\ref{table:y34J}. As can be seen from the tables the NLO coefficients are
large, which is again reflected in the large corrections shown in
Fig.~{\ref{fig:Kfac}. The K factor for the JADE scheme is
roughly 1.8-1.9, but is slightly smaller for the Geneva algorithm, typically
in the region 1.4-1.6.  

\begin{table}[h]
\begin{center}
\begin{tabular}{|c|c|c|}
\hline
$y_4^G$ & $B_{y_4^G}$ & $C_{y_4^G}$ \\ 
\hline \hline
0.0250&$( 8.00 \pm  0.04) \cdot 10^{ 2}$&$( 9.93 \pm  0.34) \cdot 10^{ 3}$  \\
0.0350&$( 5.59 \pm  0.04) \cdot 10^{ 2}$&$( 9.91 \pm  0.30) \cdot 10^{ 3}$  \\
0.0450&$( 4.15 \pm  0.03) \cdot 10^{ 2}$&$( 8.57 \pm  0.13) \cdot 10^{ 3}$  \\
0.0550&$( 3.15 \pm  0.03) \cdot 10^{ 2}$&$( 7.31 \pm  0.18) \cdot 10^{ 3}$  \\
0.0650&$( 2.47 \pm  0.02) \cdot 10^{ 2}$&$( 5.96 \pm  0.12) \cdot 10^{ 3}$  \\
0.0750&$( 1.93 \pm  0.02) \cdot 10^{ 2}$&$( 4.99 \pm  0.14) \cdot 10^{ 3}$  \\
0.0850&$( 1.50 \pm  0.02) \cdot 10^{ 2}$&$( 3.96 \pm  0.11) \cdot 10^{ 3}$  \\
0.0950&$( 1.23 \pm  0.01) \cdot 10^{ 2}$&$( 3.36 \pm  0.13) \cdot 10^{ 3}$  \\
0.1050&$( 9.88 \pm  0.12) \cdot 10^{ 1}$&$( 2.84 \pm  0.06) \cdot 10^{ 3}$  \\
0.1150&$( 7.90 \pm  0.09) \cdot 10^{ 1}$&$( 2.19 \pm  0.09) \cdot 10^{ 3}$  \\
0.1250&$( 6.07 \pm  0.08) \cdot 10^{ 1}$&$( 1.69 \pm  0.11) \cdot 10^{ 3}$  \\
0.1350&$( 4.79 \pm  0.07) \cdot 10^{ 1}$&$( 1.53 \pm  0.08) \cdot 10^{ 3}$  \\
0.1450&$( 3.84 \pm  0.06) \cdot 10^{ 1}$&$( 1.15 \pm  0.04) \cdot 10^{ 3}$  \\
0.1550&$( 3.00 \pm  0.05) \cdot 10^{ 1}$&$( 8.41 \pm  0.53) \cdot 10^{ 2}$  \\
0.1650&$( 2.26 \pm  0.04) \cdot 10^{ 1}$&$( 6.52 \pm  0.36) \cdot 10^{ 2}$  \\
0.1750&$( 1.61 \pm  0.02) \cdot 10^{ 1}$&$( 4.99 \pm  0.33) \cdot 10^{ 2}$  \\
0.1850&$( 1.21 \pm  0.02) \cdot 10^{ 1}$&$( 3.60 \pm  0.23) \cdot 10^{ 2}$  \\
0.1950&$( 8.71 \pm  0.27) \cdot 10^{ 0}$&$( 2.53 \pm  0.20) \cdot 10^{ 2}$  \\
0.2050&$( 5.70 \pm  0.16) \cdot 10^{ 0}$&$( 1.78 \pm  0.17) \cdot 10^{ 2}$  \\
0.2150&$( 3.89 \pm  0.09) \cdot 10^{ 0}$&$( 1.20 \pm  0.11) \cdot 10^{ 2}$  \\
0.2250&$( 2.41 \pm  0.06) \cdot 10^{ 0}$&$( 6.83 \pm  0.87) \cdot 10^{ 1}$  \\
0.2350&$( 1.43 \pm  0.05) \cdot 10^{ 0}$&$( 4.87 \pm  0.36) \cdot 10^{ 1}$  \\
0.2450&$( 7.69 \pm  0.30) \cdot 10^{-1}$&$( 2.57 \pm  0.25) \cdot 10^{ 1}$  \\
0.2550&$( 3.78 \pm  0.09) \cdot 10^{-1}$&$( 1.18 \pm  0.13) \cdot 10^{ 1}$  \\
0.2650&$( 1.50 \pm  0.04) \cdot 10^{-1}$&$( 4.57 \pm  0.79) \cdot 10^{ 0}$  \\
0.2750&$( 4.20 \pm  0.17) \cdot 10^{-2}$&$( 1.15 \pm  0.35) \cdot 10^{ 0}$  \\
0.2850&$( 4.59 \pm  0.39) \cdot 10^{-3}$&$( 1.16 \pm  0.65) \cdot 10^{-1}$  \\
0.2950&$( 5.37 \pm  0.91) \cdot 10^{-5}$&$( 2.15 \pm  1.11) \cdot 10^{-3}$  \\
\hline
\end{tabular}
\caption{The leading and next-to-leading order coefficients for the jet
transition variable in the Geneva-E algorithm $y_4^G$.}
\label{table:y34G}
\end{center}
\end{table}

\begin{table}[h]
\begin{center}
\begin{tabular}{|c|c|c|}
\hline
$y_4^J$ & $B_{y_4^J}$ & $C_{y_4^J}$ \\ 
\hline \hline
0.0075&$( 6.02 \pm  0.01) \cdot 10^{ 2}$&$( 1.75 \pm  0.01) \cdot 10^{ 4}$  \\
0.0125&$( 3.60 \pm  0.01) \cdot 10^{ 2}$&$( 1.33 \pm  0.02) \cdot 10^{ 4}$  \\
0.0175&$( 2.47 \pm  0.01) \cdot 10^{ 2}$&$( 1.02 \pm  0.04) \cdot 10^{ 4}$  \\
0.0225&$( 1.78 \pm  0.01) \cdot 10^{ 2}$&$( 7.63 \pm  0.32) \cdot 10^{ 3}$  \\
0.0275&$( 1.34 \pm  0.01) \cdot 10^{ 2}$&$( 6.19 \pm  0.16) \cdot 10^{ 3}$  \\
0.0325&$( 1.01 \pm  0.01) \cdot 10^{ 2}$&$( 4.76 \pm  0.12) \cdot 10^{ 3}$  \\
0.0375&$( 7.88 \pm  0.08) \cdot 10^{ 1}$&$( 3.86 \pm  0.11) \cdot 10^{ 3}$  \\
0.0425&$( 6.19 \pm  0.05) \cdot 10^{ 1}$&$( 3.07 \pm  0.16) \cdot 10^{ 3}$  \\
0.0475&$( 4.99 \pm  0.05) \cdot 10^{ 1}$&$( 2.38 \pm  0.12) \cdot 10^{ 3}$  \\
0.0525&$( 3.89 \pm  0.05) \cdot 10^{ 1}$&$( 2.08 \pm  0.11) \cdot 10^{ 3}$  \\
0.0575&$( 3.13 \pm  0.05) \cdot 10^{ 1}$&$( 1.54 \pm  0.05) \cdot 10^{ 3}$  \\
0.0625&$( 2.43 \pm  0.04) \cdot 10^{ 1}$&$( 1.26 \pm  0.03) \cdot 10^{ 3}$  \\
0.0675&$( 1.90 \pm  0.03) \cdot 10^{ 1}$&$( 9.68 \pm  0.58) \cdot 10^{ 2}$  \\
0.0725&$( 1.49 \pm  0.04) \cdot 10^{ 1}$&$( 7.70 \pm  0.35) \cdot 10^{ 2}$  \\
0.0775&$( 1.21 \pm  0.02) \cdot 10^{ 1}$&$( 5.89 \pm  0.41) \cdot 10^{ 2}$  \\
0.0825&$( 9.38 \pm  0.18) \cdot 10^{ 0}$&$( 4.83 \pm  0.35) \cdot 10^{ 2}$  \\
0.0875&$( 6.94 \pm  0.09) \cdot 10^{ 0}$&$( 3.50 \pm  0.19) \cdot 10^{ 2}$  \\
0.0925&$( 5.36 \pm  0.11) \cdot 10^{ 0}$&$( 2.48 \pm  0.27) \cdot 10^{ 2}$  \\
0.0975&$( 3.85 \pm  0.06) \cdot 10^{ 0}$&$( 1.93 \pm  0.19) \cdot 10^{ 2}$  \\
0.1025&$( 2.84 \pm  0.07) \cdot 10^{ 0}$&$( 1.26 \pm  0.11) \cdot 10^{ 2}$  \\
0.1075&$( 1.97 \pm  0.07) \cdot 10^{ 0}$&$( 9.99 \pm  1.22) \cdot 10^{ 1}$  \\
0.1125&$( 1.30 \pm  0.06) \cdot 10^{ 0}$&$( 6.69 \pm  0.94) \cdot 10^{ 1}$  \\
0.1175&$( 8.32 \pm  0.37) \cdot 10^{-1}$&$( 3.57 \pm  0.52) \cdot 10^{ 1}$  \\
0.1225&$( 4.94 \pm  0.07) \cdot 10^{-1}$&$( 2.36 \pm  0.44) \cdot 10^{ 1}$  \\
0.1275&$( 3.05 \pm  0.10) \cdot 10^{-1}$&$( 1.85 \pm  0.38) \cdot 10^{ 1}$  \\
0.1325&$( 1.70 \pm  0.03) \cdot 10^{-1}$&$( 8.38 \pm  3.15) \cdot 10^{ 0}$  \\
0.1375&$( 8.94 \pm  0.29) \cdot 10^{-2}$&$( 4.99 \pm  1.15) \cdot 10^{ 0}$  \\
0.1425&$( 4.20 \pm  0.12) \cdot 10^{-2}$&$( 2.01 \pm  0.38) \cdot 10^{ 0}$  \\
0.1475&$( 1.67 \pm  0.07) \cdot 10^{-2}$&$( 1.08 \pm  0.73) \cdot 10^{ 0}$  \\
0.1525&$( 5.51 \pm  0.44) \cdot 10^{-3}$&$( 3.94 \pm  2.32) \cdot 10^{-1}$  \\
0.1575&$( 8.48 \pm  0.58) \cdot 10^{-4}$&$( 4.37 \pm  2.44) \cdot 10^{-2}$  \\
\hline
\end{tabular}
\caption{The leading and next-to-leading order coefficients for the jet
transition variable in the JADE-E0 algorithm $y_4^J$.}
\label{table:y34J}
\end{center}
\end{table}

\clearpage
\section{Comparison with experimental data}
\setcounter{equation}{0}
\label{sec:data}

Four jet event shape observables have been studied extensively by the four LEP
experiments.   However, the most complete analysis of event shape variables has
been carried out by the DELPHI collaboration  \cite{4jetdata}. Here, they study
all of the event shape variables discussed in section~\ref{sec:shape}. 
Distributions based on charged particles alone as well as
charged and neutral particles are presented. In this section, we wish to
examine whether or not these event shapes can be described by fixed order
perturbation theory. As discussed earlier, to avoid numerical instabilities in
the  infrared region where fixed order perturbation theory is no longer valid
we impose a cut on the smallness of the variable that is generally equal to the
lower edge of the second bin. More precisely, that is, 
\begin{eqnarray} 
D             & > & 0.008,\nonumber \\ 
T_{\rm minor} & > & 0.02,\nonumber \\ 
M_L^2/s       & > & 0.01,\nonumber \\ 
B_{\min}      & > & 0.01,\nonumber \\ 
A             & > & 0.005, \nonumber \\
y_4^D         & > & 0.002,\nonumber \\ 
y_4^J         & > & 0.002. 
\label{eq:avval}
\end{eqnarray} 

The experimental distributions are normalised to the hadronic cross section
(rather than the Born cross section) and are also not weighted by the
observable, but are rather,
\begin{equation} \frac{1}{\sigma_{\rm had}}
\cdot  \, \frac{d\sigma}{dO_4} = \left( \frac{\alpha_s(\mu)}{2\pi} \right)^2
\frac{B_{O_4}}{O_4} +\left( \frac{\alpha_s(\mu)}{2\pi} \right)^3 \left( 2
\beta_0 \log \left(\frac{\mu^2}{s}\right)\frac{ B_{O_4}}{O_4} +\frac{
C_{O_4}-2B_{O_4}}{O_4} \right).
\label{eq:dist}
\end{equation}
Throughout, we choose $\alpha_s(M_Z) = 0.118$ which is consistent with the
current world average \cite{worldav}. In each case, the theoretical predictions
have been evaluated using bins of the same size as in the experiment and
therefore appear as histograms in the plots. The data is corrected for detector
effects, but not for hadronisation effects.

\begin{figure}[ht]
\begin{center}
(a)
\psfig{figure=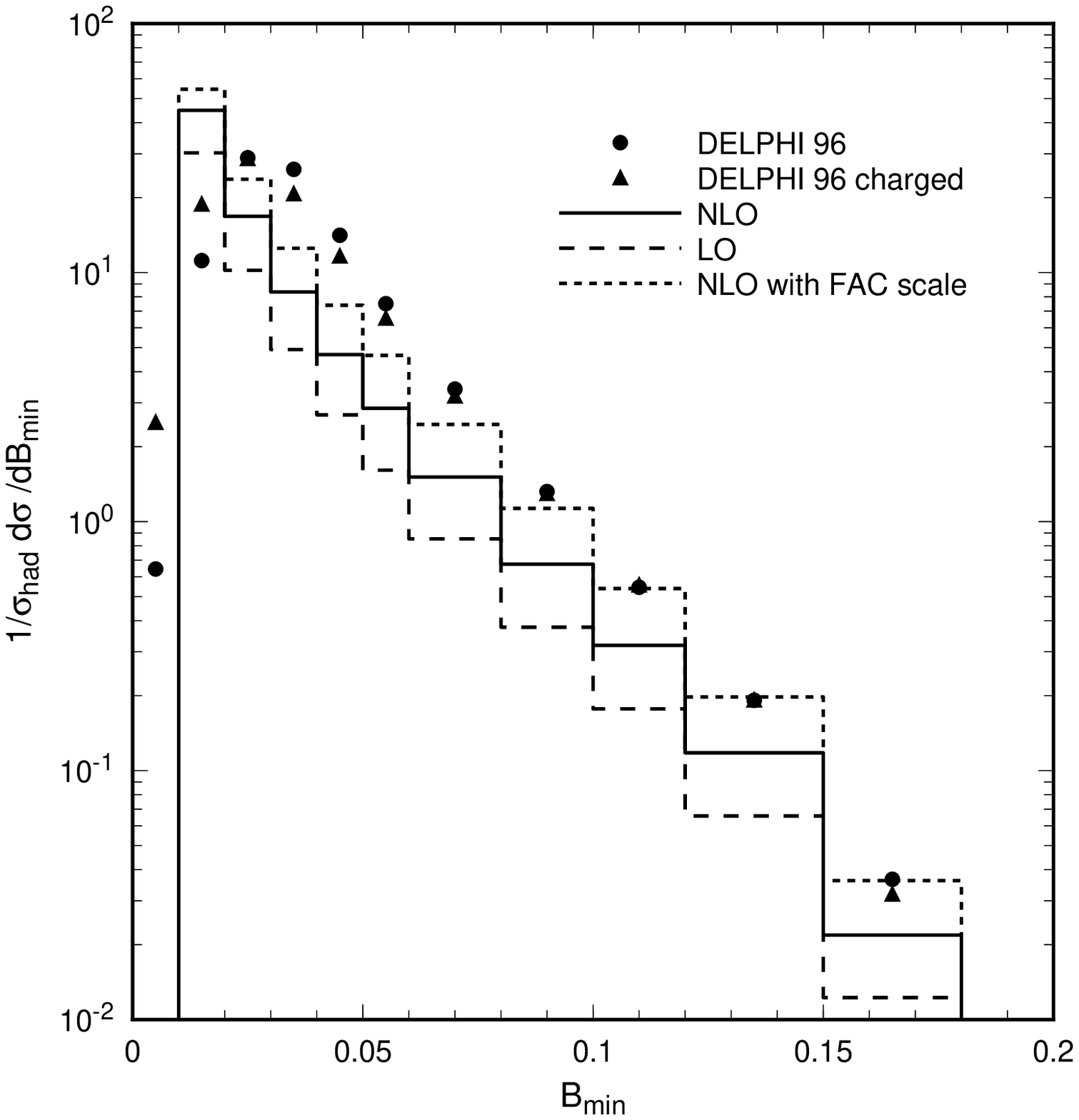,width=7cm}
(b)
\psfig{figure=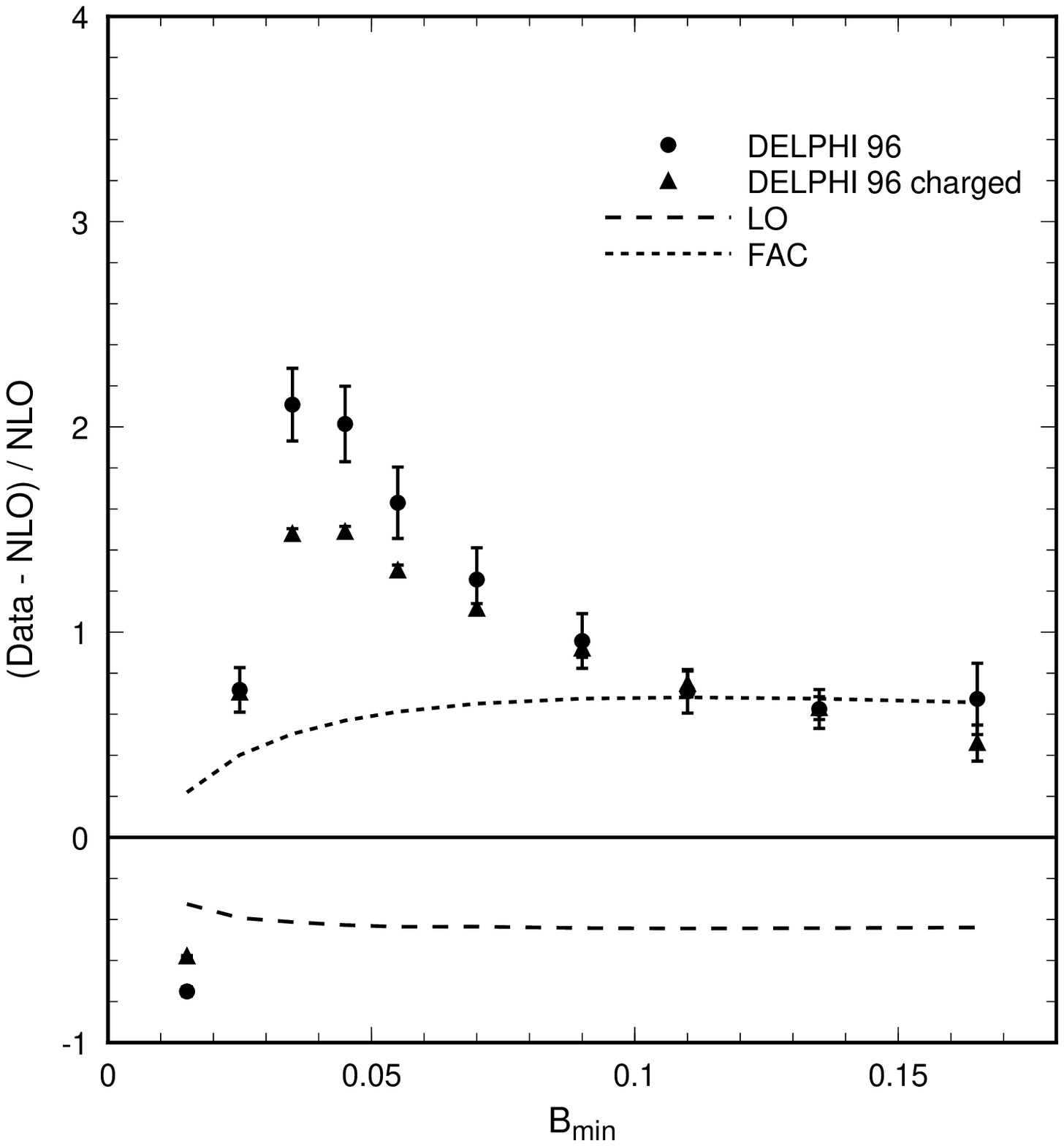,width=7cm}
\end{center}
\caption[]{The leading order (dashed) and next-to-leading order (solid)
predictions evaluated at the physical scale $\mu = \sqrt{s} = M_Z$
for (a) $1/\sigma_{\rm had}  \cdot d\sigma/dB_{\rm min}$ 
compared to the published DELPHI data \cite{4jetdata} and
(b) the difference between data and NLO theory (normalised to NLO).
The short-dashed line shows the next-to-leading order prediction using the
FAC scale (see eq.~(\ref{eq:facscale})).}
\label{fig:Bdata}
\end{figure}

\begin{figure}[ht]
\begin{center}
(a)
\psfig{figure=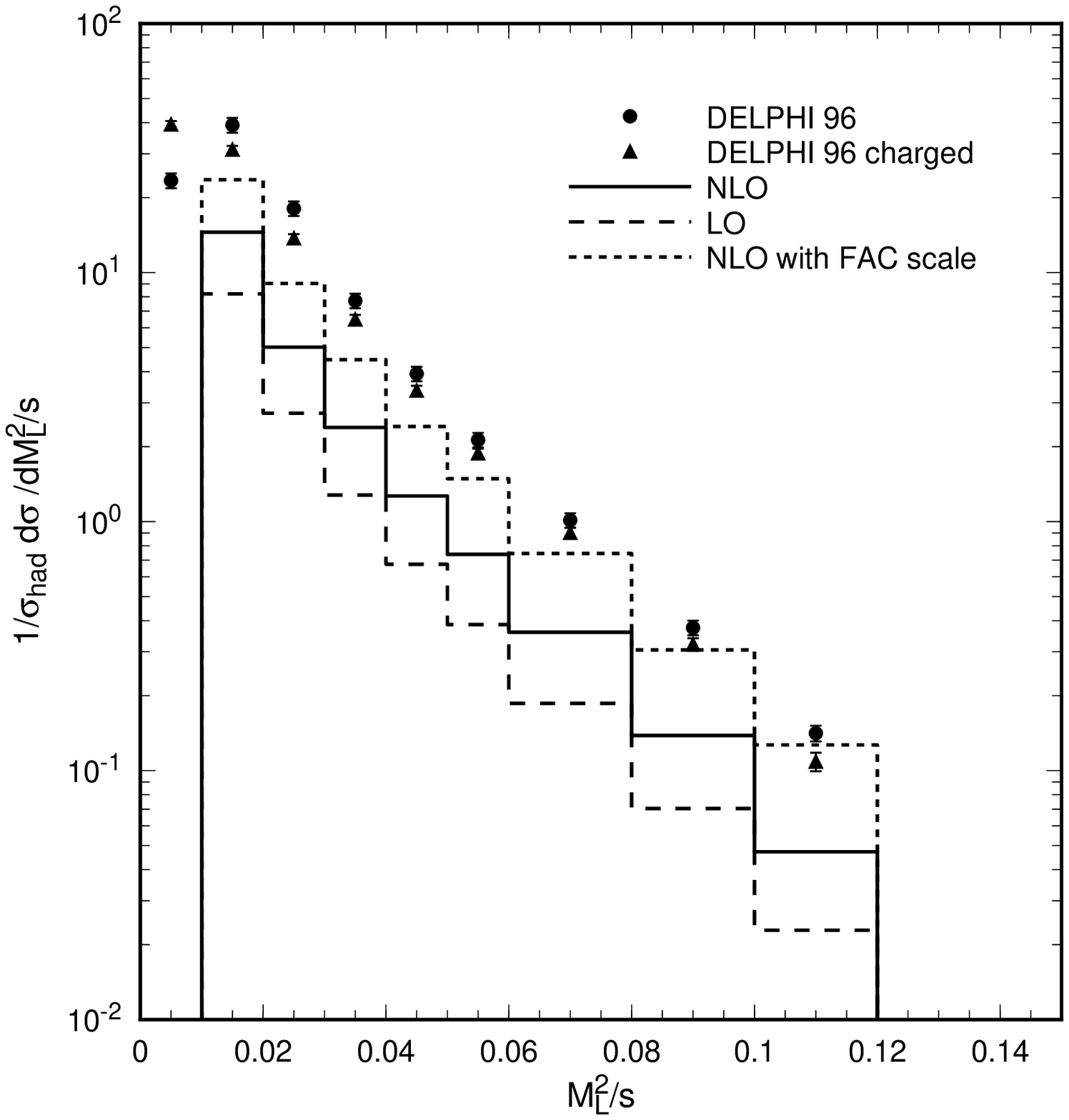,width=7cm}
(b)
\psfig{figure=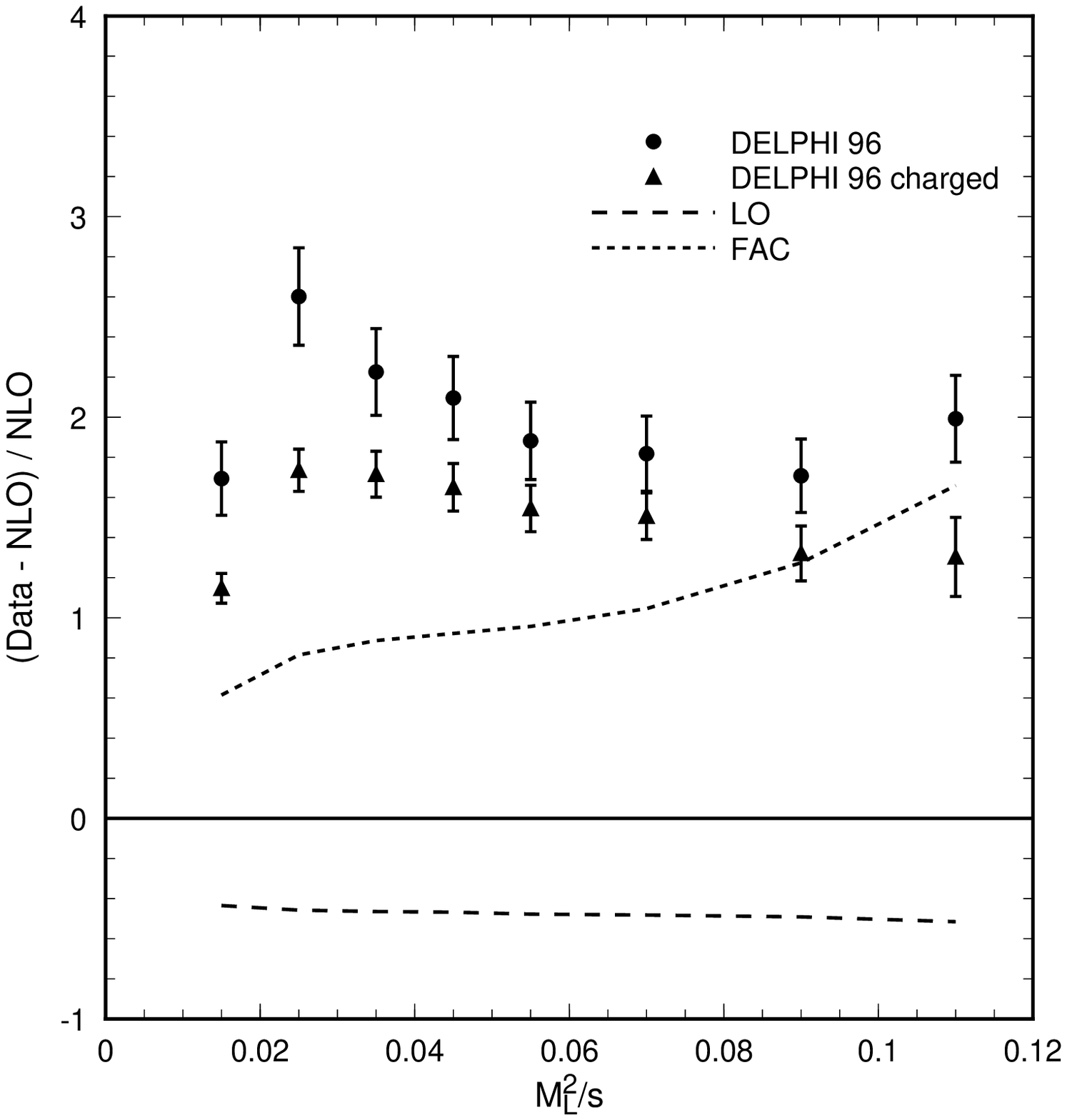,width=7cm}
\end{center}
\caption[]{The leading order (dashed) and next-to-leading order (solid)
predictions evaluated at the physical scale $\mu = \sqrt{s} = M_Z$
for
(a) $1/\sigma_{\rm had}  \cdot d\sigma/d(M_L^2/s)$ 
compared to the published DELPHI data \cite{4jetdata}
and (b) the difference between data and NLO theory (normalised to NLO).
The short-dashed line shows the next-to-leading order prediction using the
FAC scale (see eq.~(\ref{eq:facscale})).}
\label{fig:Ldata}
\end{figure}

\begin{figure}[tbp]
\begin{center}
(a)
\psfig{figure=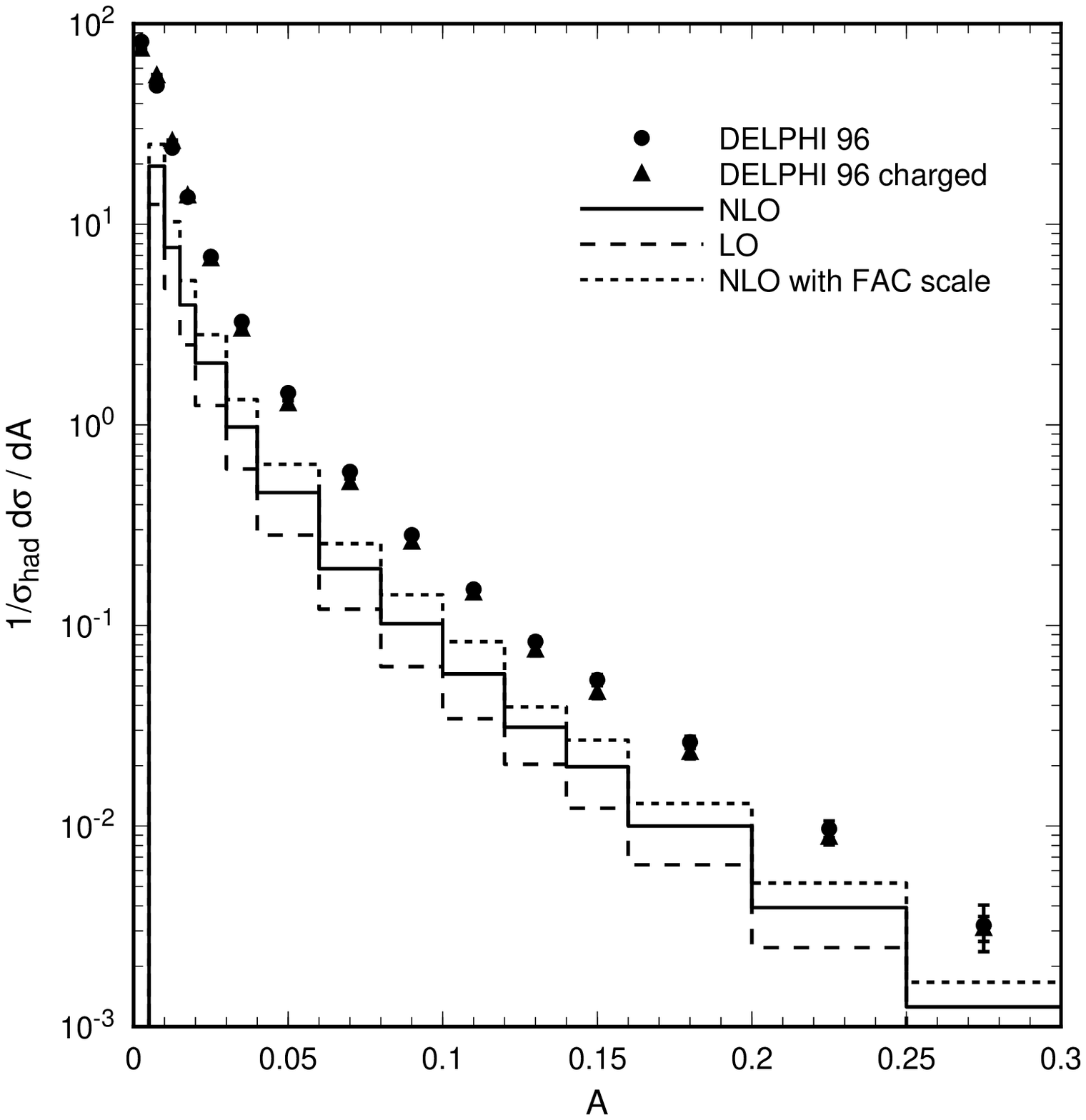,width=7cm}
(b)
\psfig{figure=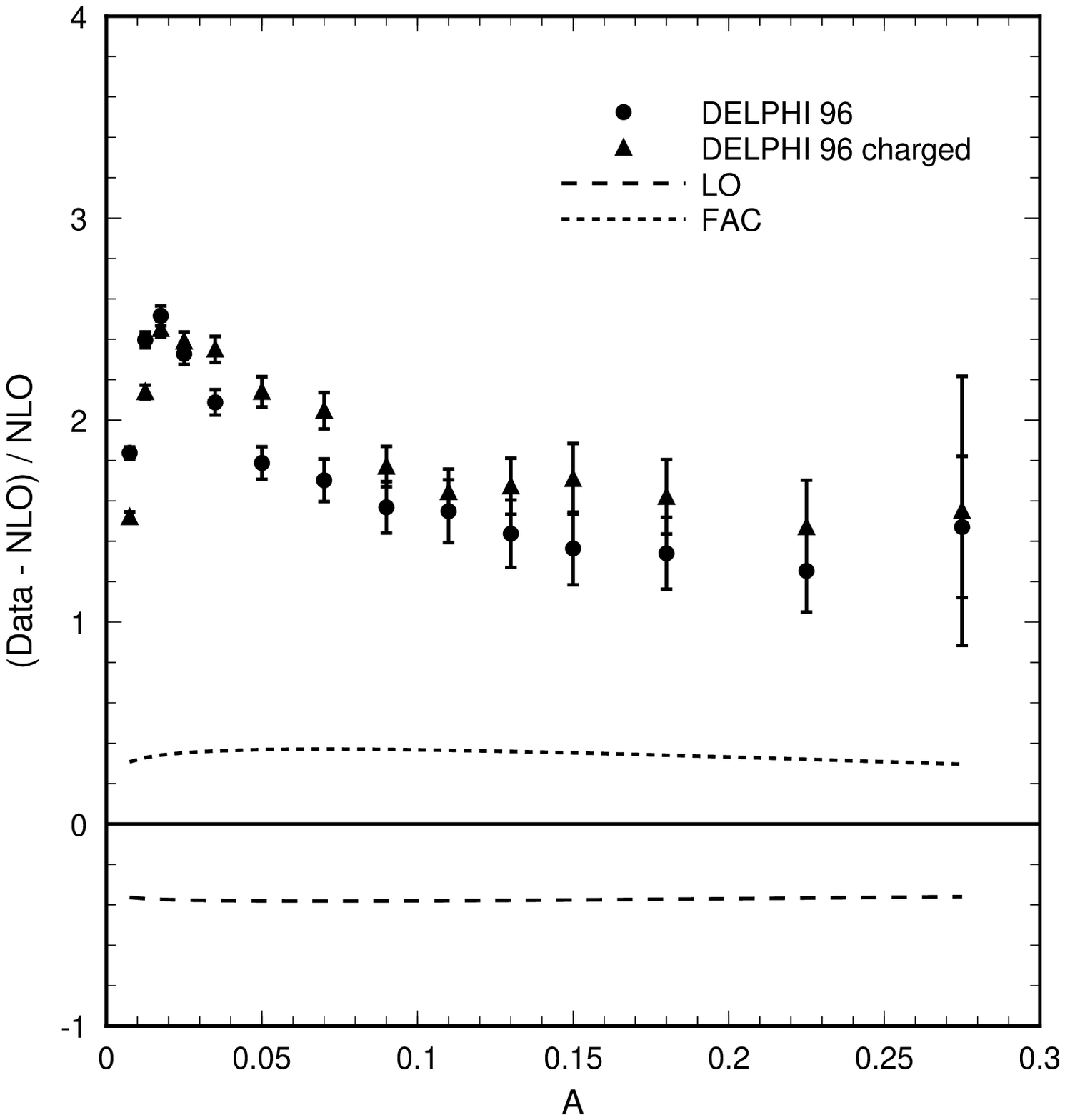,width=7cm}
\end{center}
\caption[]{The leading order (dashed) and next-to-leading order (solid)
predictions evaluated at the physical scale $\mu = \sqrt{s} = M_Z$
for (a) $1/\sigma_{\rm had}  \cdot d\sigma/dA$ 
compared to the published DELPHI data \cite{4jetdata}
and (b) the difference between data and NLO theory (normalised to NLO).
The short-dashed line shows the next-to-leading order prediction using the
FAC scale (see eq.~(\ref{eq:facscale})).}
\label{fig:Adata}
\end{figure}

Figs.~\ref{fig:Bdata}, \ref{fig:Ldata} and \ref{fig:Adata} show the comparison
between the leading order and next-to-leading order  predictions evaluated at
the physical scale $\mu = \sqrt{s} = M_Z$  for narrow jet broadening, light
hemisphere mass and Aplanarity with the published DELPHI data \cite{4jetdata}.
We see that in all three cases, the LO prediction undershoots the data by a
significant factor (about a factor of four), and that including the NLO
correction improves the situation but still gives a rate that is much lower
than the data.     However,  the NLO prediction still contains a large
renormalisation scale  uncertainty. Usually, one varies the choice of scale
about the physical scale by a factor of two or so, but as discussed earlier,
the FAC scale defined in eq.~(\ref{eq:facscale}) is an attractive alternative
choice in that the known ultraviolet logarithms are resummed \cite{corgi}. For
both of these variables, the FAC scale is significantly less than the physical
scale, for example, for $B_{\rm min}$ , $\mu^{FAC} \sim 0.06 \sqrt{s}$. This
has the effect of increasing $\alpha_s$, thereby increasing the NLO prediction
and in both cases, we see much improved agreement at larger values of the
observable. At smaller values, and particularly in the region where the data
turns over the  agreement is still poor.    This, of course, is where the
infrared logarithms are large and need to be resummed. Furthermore, we also
expect non-perturbative hadronisation effects or power corrections to influence
the perturbative shape of the distribution we have been concerned with
\cite{pc,thrust}.  These contributions (together with resummation of the
infrared logarithms) have played an important role in extracting useful
information from analyses of three jet shape variables, and are likely to be
important in analysing four jet event shapes.

\begin{figure}[tbp]
\begin{center}
(a)
\psfig{figure=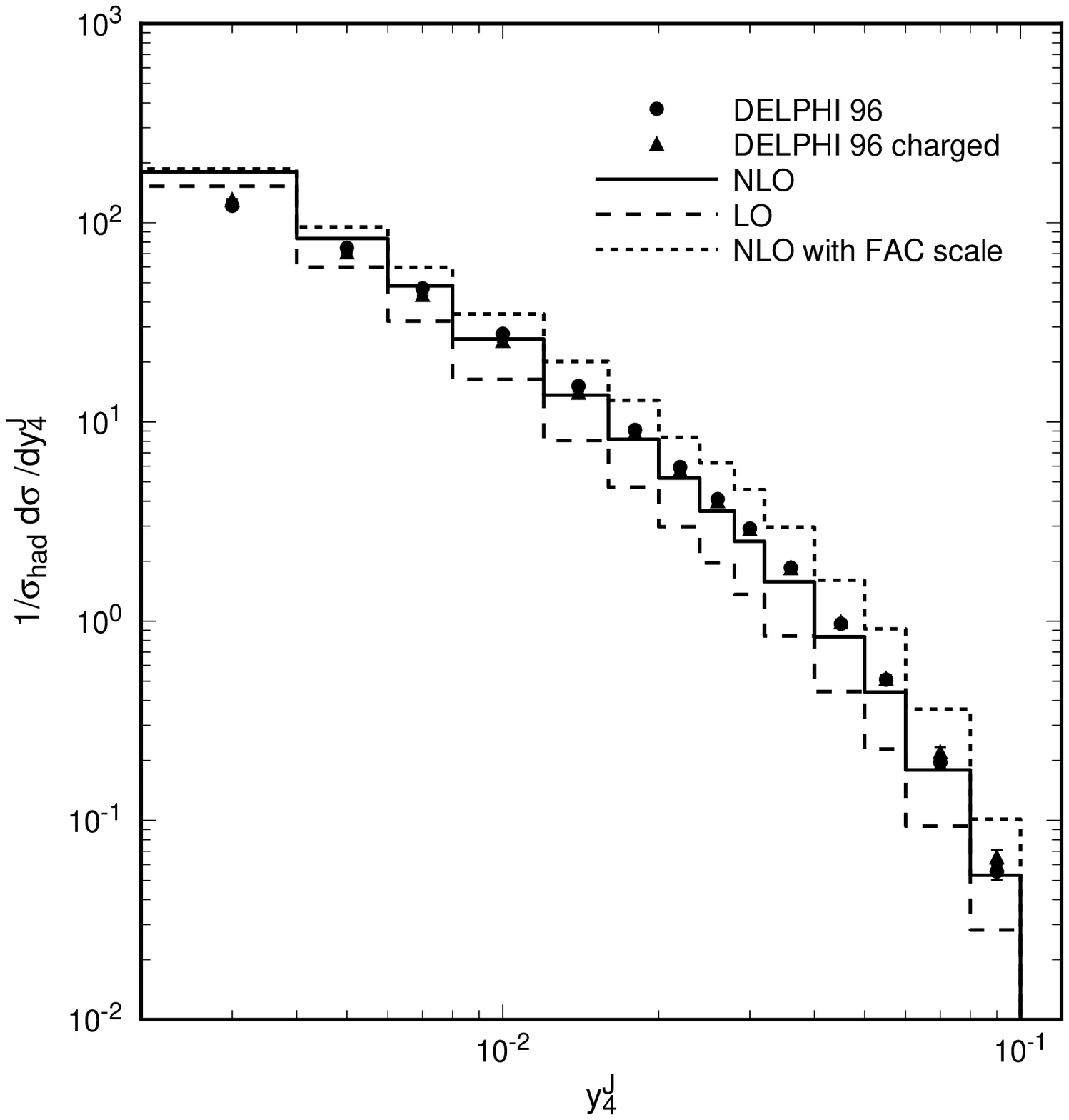,width=7cm}
(b)
\psfig{figure=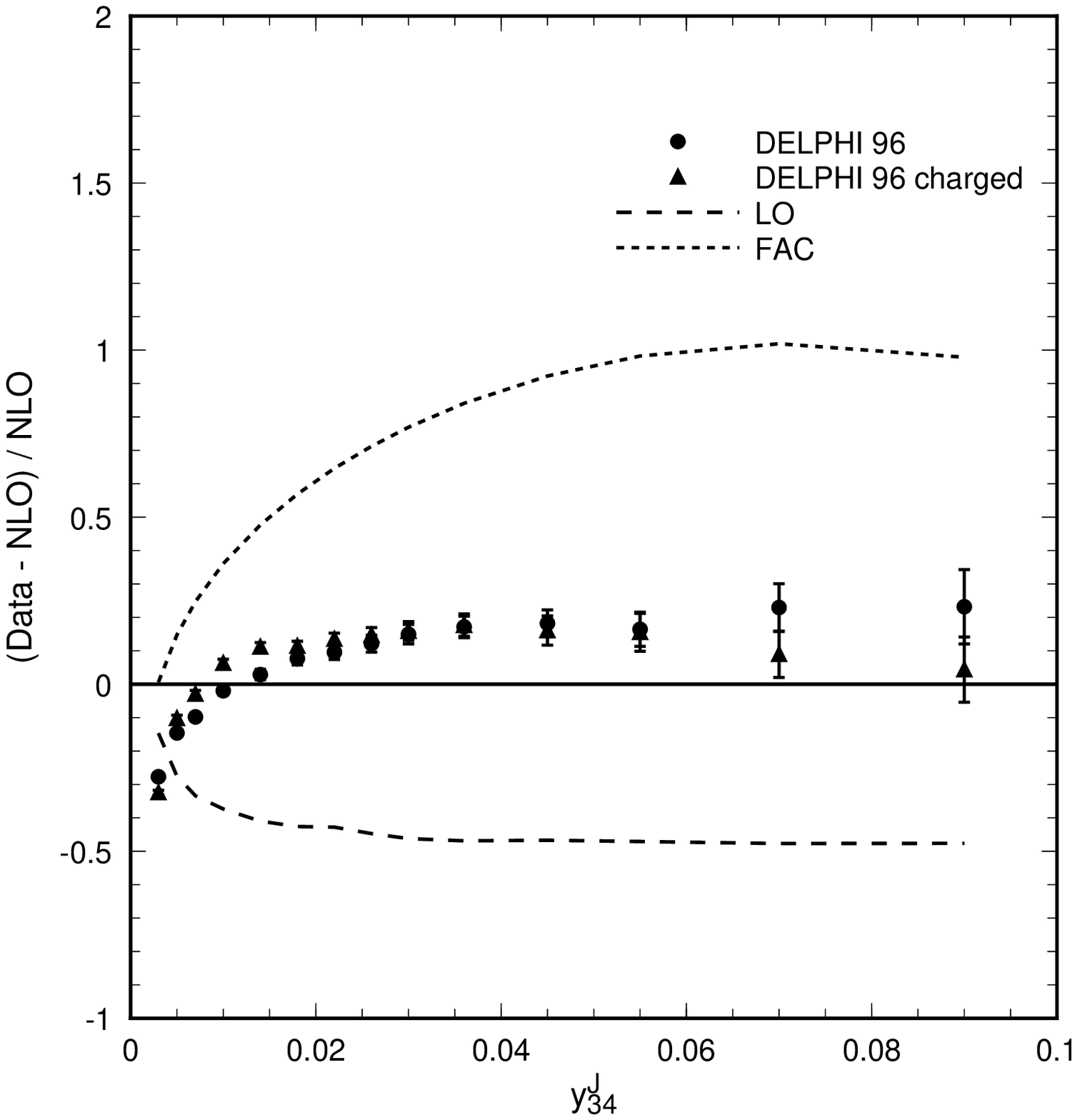,width=7cm}
\end{center}
\caption[]{The leading order (dashed) and next-to-leading order (solid)
predictions evaluated at the physical scale $\mu = \sqrt{s} = M_Z$
for (a) $1/\sigma_{\rm had}  \cdot d\sigma/dy_4^J$ 
compared to the published DELPHI data \cite{4jetdata}
and (b) the difference between data and NLO theory (normalised to NLO).
The short-dashed line shows the next-to-leading order prediction using the
FAC scale (see eq.~(\ref{eq:facscale})).}
\label{fig:Jdata}
\end{figure}

\begin{figure}[tbp]
\begin{center}
(a)
\psfig{figure=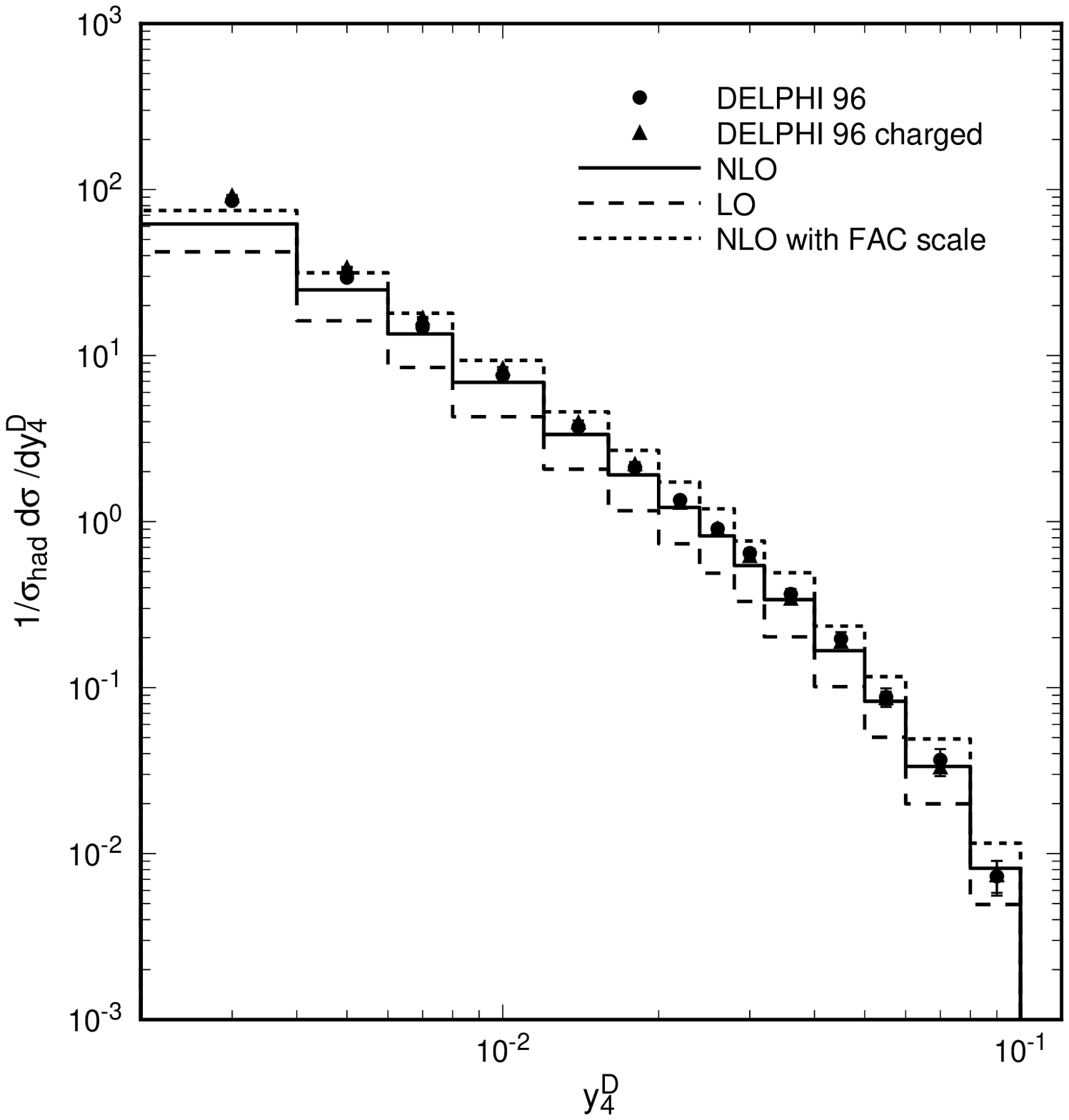,width=7cm}
(b)
\psfig{figure=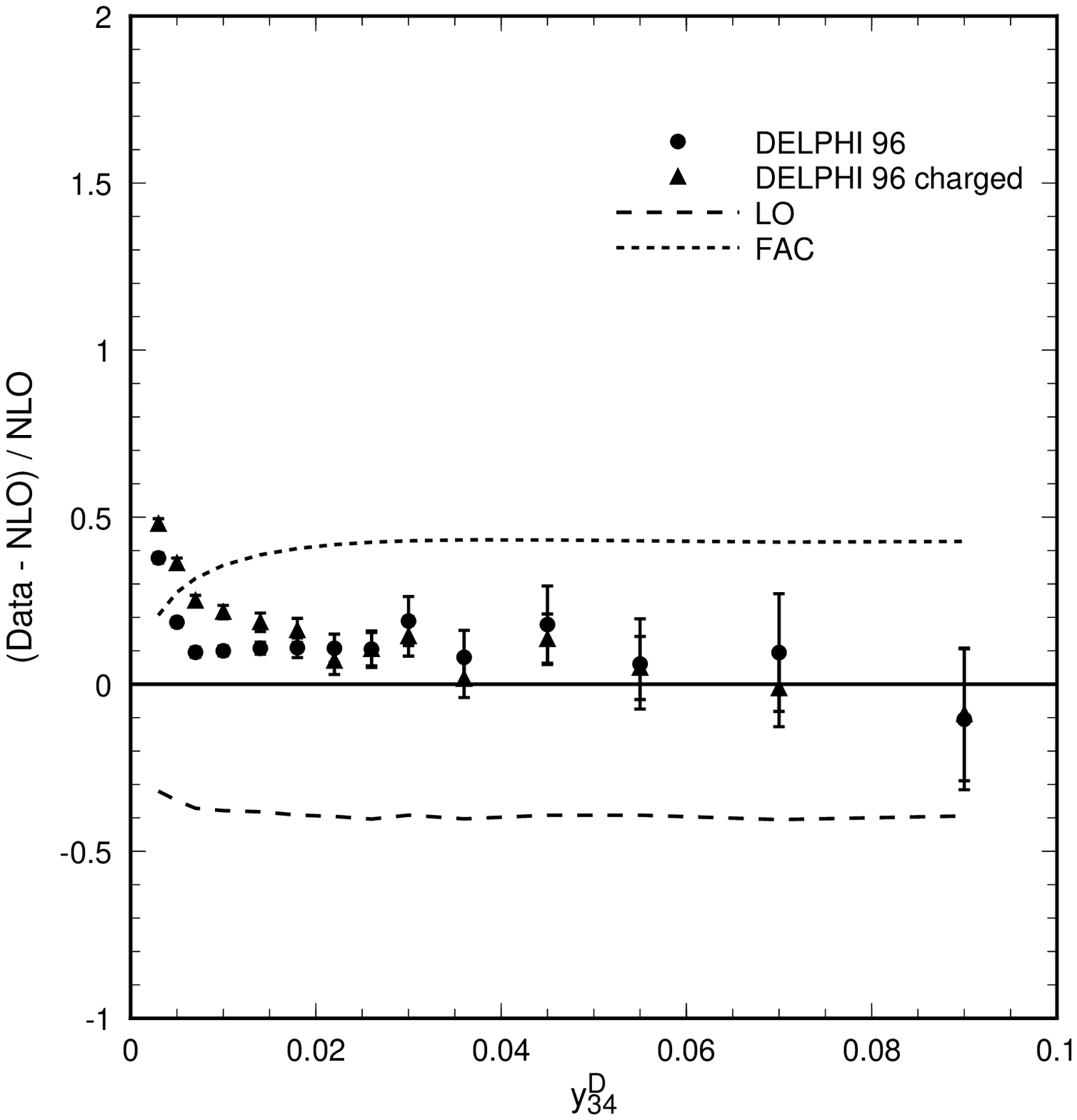,width=7cm}
\end{center}
\caption[]{The leading order (dashed) and next-to-leading order (solid)
predictions evaluated at the physical scale $\mu = \sqrt{s} = M_Z$
for (a) $1/\sigma_{\rm had}  \cdot d\sigma/dy_4^D$ 
compared to the published DELPHI data \cite{4jetdata}
and (b) the difference between data and NLO theory (normalised to NLO).
The short-dashed line shows the next-to-leading order prediction using the
FAC scale (see eq.~(\ref{eq:facscale})).}
\label{fig:Kdata}
\end{figure}

A similar comparison of the perturbative predictions for the jet transition
rates with the DELPHI measurements\footnote{The DELPHI data gives the
differential jet rate rather than the jet transition variable.  Up to a small
($\sim $ few \%) correction from five (and more) jet events falling into a four
jet configuration, the two quantities coincide at fixed order.} is made in
Figs.~\ref{fig:Jdata} and \ref{fig:Kdata}. Once again, the LO distribution lies
well below the data.  This time, the NLO prediction lies much closer to the
data for most of the available kinematic range.  The FAC scale rate usually
lies above the NLO prediction so that the data lies within the range of
uncertainty engendered by the renormalisation group. 

For completeness, Figs.~\ref{fig:Ddata} and \ref{fig:Tdata} show the DELPHI
data and perturbative predictions for the $D$ parameter and $T_{\rm minor}$
repectively.   As expected from the analysis of Nagy and Tr{\'o}cs{\'a}nyi
\cite{debrecen},  the LO prediction for $D$ is too low by a factor of about
four, while at the physical scale $\mu = \sqrt{s}=M_Z$ the NLO distribution is
roughly twice as large but still lies a factor of two below the data. 
However,  for the FAC scale (which for the $D$ parameter is approximately
$0.035\sqrt{s}$)  the prediction overshoots by 50\% or so for $D > 0.1$ where
the fixed order prediction is least affected by large infrared logs.

The importance of resumming these logs is clearly shown in Fig.~\ref{fig:Tdata}
where the $T_{\rm minor}$ distribution is shown. For $T_{\rm minor}>0.1$ the
data again lies between the next-to-leading order predictions at the physical
and FAC scales (which encompass an uncertainty of about a factor of almost
three for $T_{\rm minor} \sim 0.2$). However, the turn-over at $T_{\rm
minor}=0.1$ cannot be modelled without resumming the large logs which cause the
perturbative prediction to grow rapidly. The same is true at small values of
the light hemisphere mass and narrow jet broadening although there the effects
are less pronounced because of the choice of bin sizes.

\begin{figure}[t]
\begin{center}
(a)
\psfig{figure=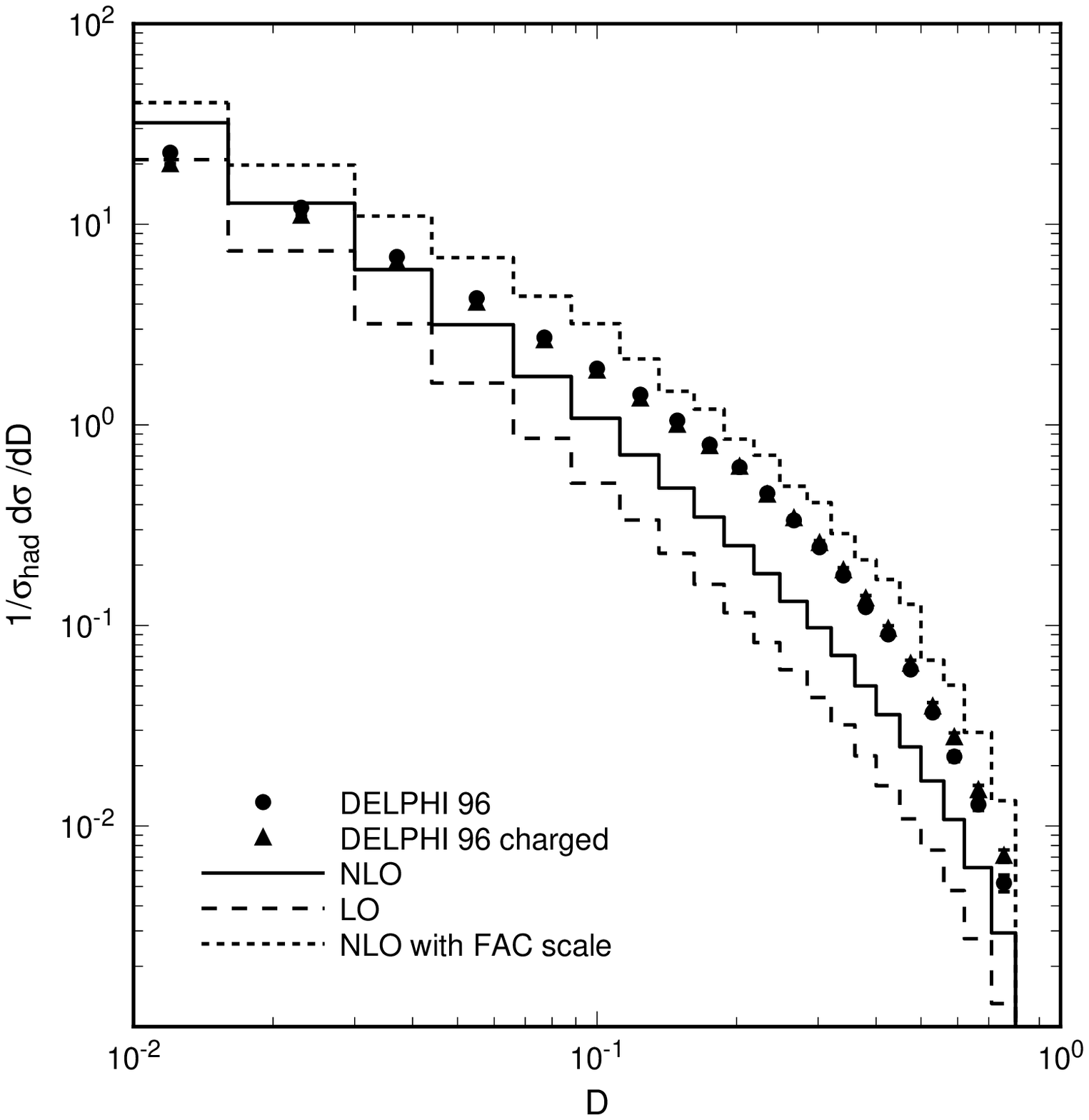,width=7cm}
(b)
\psfig{figure=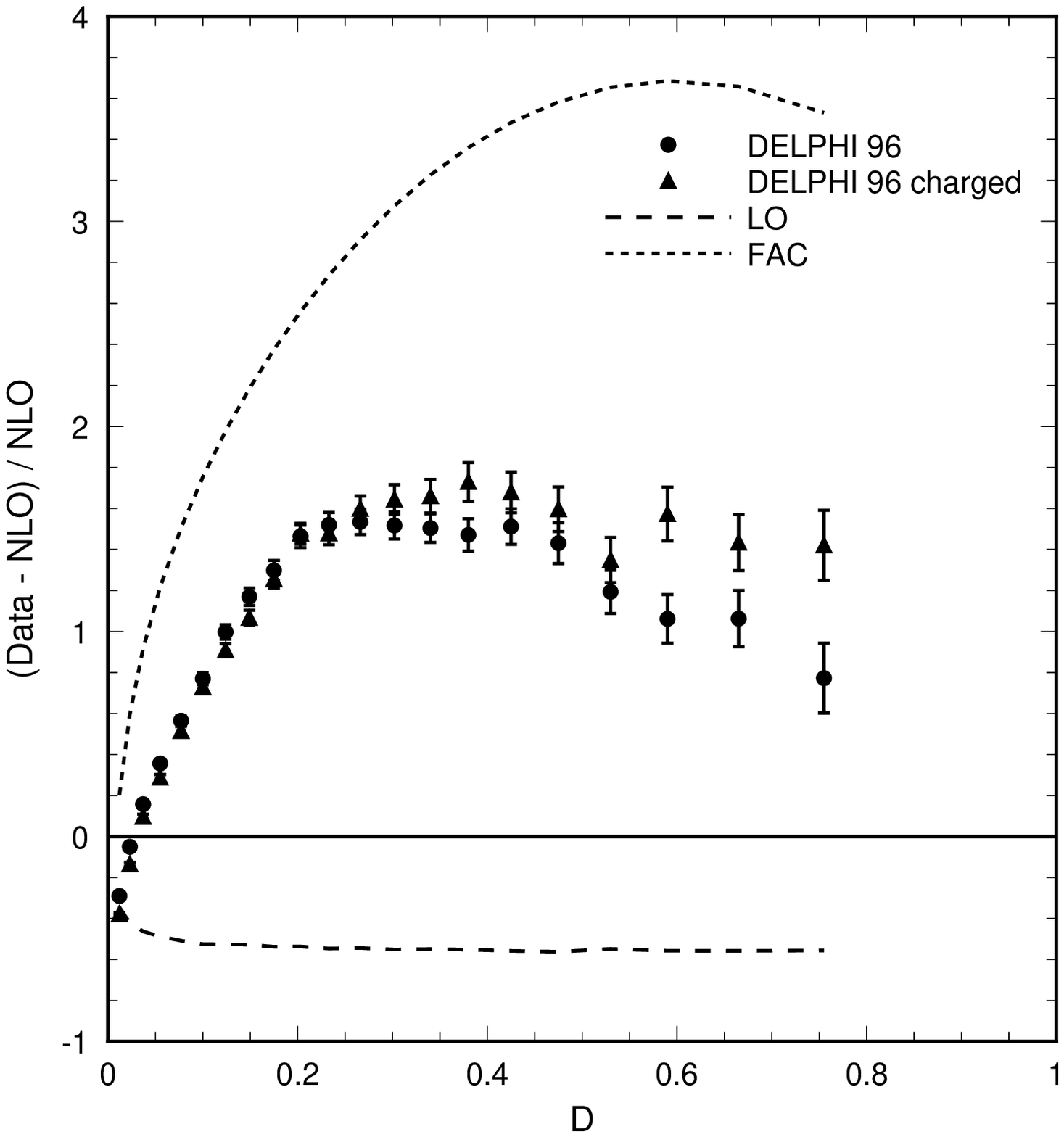,width=7cm}
\end{center}
\caption[]{The leading order (dashed) and next-to-leading order (solid)
predictions evaluated at the physical scale $\mu = \sqrt{s} = M_Z$
for (a) $1/\sigma_{\rm had}  \cdot d\sigma/dD$ 
compared to the published DELPHI data \cite{4jetdata}
and (b) the difference between data and NLO theory (normalised to NLO).
The short-dashed line shows the next-to-leading order prediction using the
FAC scale (see eq.~(\ref{eq:facscale})).}
\label{fig:Ddata}
\end{figure}

\begin{figure}[t]
\begin{center}
(a)
\psfig{figure=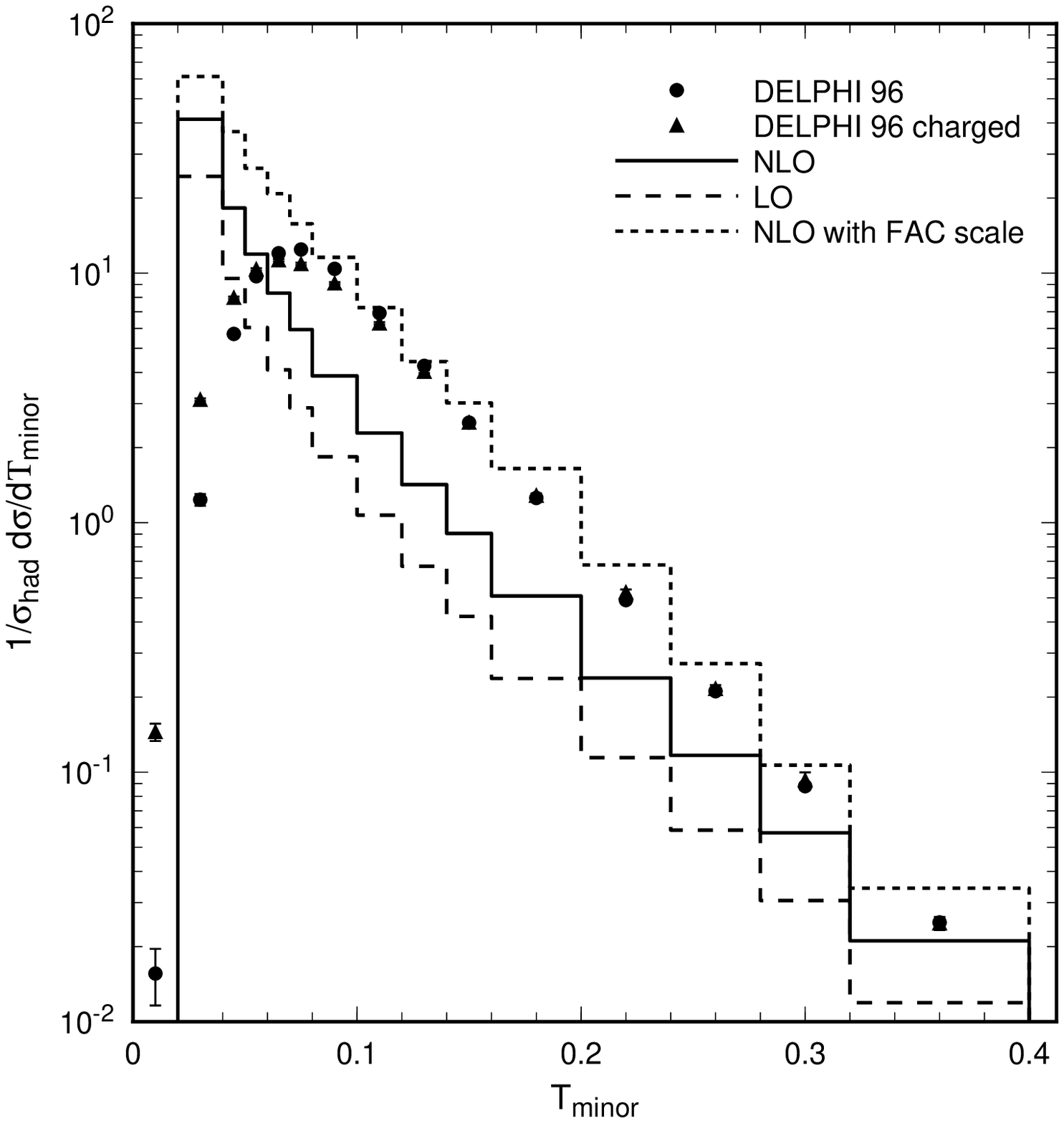,width=7cm}
(b)
\psfig{figure=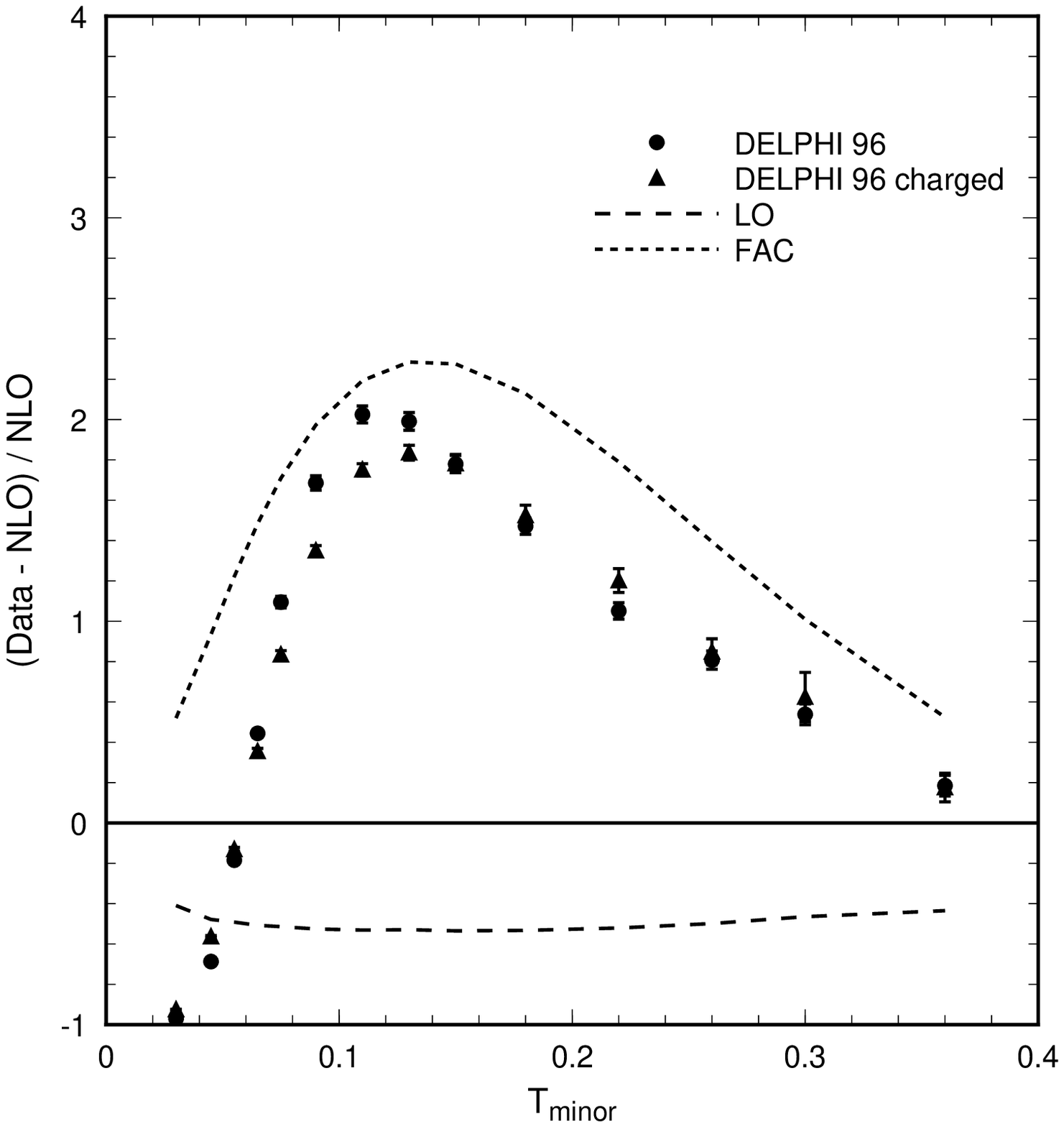,width=7cm}
\end{center}
\caption[]{The leading order (dashed) and next-to-leading order (solid)
predictions evaluated at the physical scale $\mu = \sqrt{s} = M_Z$
for (a) $1/\sigma_{\rm had}  \cdot d\sigma/dT_{\rm minor}$ 
compared to the published DELPHI data \cite{4jetdata}
and (b) the difference between data and NLO theory (normalised to NLO).
The short-dashed line shows the next-to-leading order prediction using the
FAC scale (see eq.~(\ref{eq:facscale})).}
\label{fig:Tdata}
\end{figure}

\clearpage
\section{Conclusions}
\setcounter{equation}{0}
\label{sec:Conc}

In this paper we have introduced a new Monte Carlo program for the calculation
of 4 jet like observables in electron-positron annihilation. This program, {\tt
EERAD2}, is based on the known  squared matrix elements for $\gamma^* \to 4$
and 5 partons and numerically implements the necessary cancellations between
the different final states using a hybrid of the commonly used subtraction and
slicing schemes. This infrared cancellation scheme is detailed in the appendix.

We have checked that the numerical results obtained with {\tt EERAD2}  are
consistent with the two other available four jet programs {\tt MENLO PARC} and
{\tt DEBRECEN} by recalculating the distributions for some of the previously
determined four jet event variables (such as the $D$ parameter, thrust minor
and the jet transition rate for the Durham jet algorithm) as well as the four
jet rate.  Within the (estimated) statistical Monte Carlo errors, there is
excellent agreement.

We have also presented the leading and next-to-leading order scale independent
coefficients for some previously uncalculated observables; the light hemisphere
mass, narrow jet broadening, Aplanarity and the four jet transition variables
with respect to the JADE and Geneva jet finding algorithms. For each of these
observables, the  next-to-leading order corrections calculated at the physical
scale significantly increase the rate compared to leading order (see
fig.~\ref{fig:Kfac}). The renormalisation scale dependence is also rather
large.

Furthermore, for each of these observables, we have made a comparison with the 
published  data collected at LEP 1 energies by the DELPHI collaboration. With
the exception of the  $y_4$ distributions, the data lies well above the NLO
prediction, apart from  in the infrared region where the NLO prediction grows
rapidly (and where resummation of large logarithms is essential). Using the FAC
scale, which is of the order of $0.1$--$1$\% of the physical scale, increases
the predicted rate and in general produces a slightly better agreement with the
data. Taking together the size of the NLO corrections, the renormalisation
scale dependence and poor agreement with data it appears that the inclusion of
even higher order corrections and/or power corrections will be necessary to
extract any useful information from these observations.

Although at first sight this may seem discouraging, it is instructive to
compare this situation with the well-established results for the 3 jet-like
variable $1-{\rm Thrust}$.  In this case, the next-to-leading order coefficient
is also large compared to the leading order term, resulting in a K factor at
the physical scale which varies from $1.4$--$1.6$ throughout most of the
kinematic range of the variable. These values are very similar to the case for
the four jet observables that we have already discussed. In addition, it is
also possible to compare the pure perturbative prediction for $1-{\rm Thrust}$
with the DELPHI data, as we have done for the four jet variables in
section~\ref{sec:results}. This yields results which are qualitatively very
similar to those shown for the $D$-parameter and $T_{\rm minor}$ in
Figs.~\ref{fig:Ddata} and~\ref{fig:Tdata}. In these cases, it is clear that the
perturbative prediction -- either at the physical scale or using the FAC choice
-- is insufficient to describe the data.  However, after resummation of large
logarithms and the inclusion of non-perturbative power corrections proportional
to the inverse of the energy scale of the process, $1/Q$, it is known that the
data for $1-{\rm Thrust}$ can be described very well \cite{thrust}. In fact the thrust
distribution forms a text-book example of how to interpret hadronic final
states within QCD.

From this we conclude that discarding four jet-like event shape variables as
unreliable at next-to-leading order is premature without proper consideration
of the types of non-perturbative terms that have been successfully included for
a variety of three jet-like observables. In this we disagree with the
conclusions of reference~\cite{debrecen}.

In addition to these correction terms, all of the distributions that we have
considered exhibit the infrared logarithmic behaviour (as the observable tends
to zero) that is inevitable in a fixed order calculation. These logarithms are
present in the coefficients $B_{O_4}$ and $C_{O_4}$ and can be parametrized in
the form shown in eq.~(\ref{eq:logbehaviour}). By performing a fit to the
distributions at low enough values of the variable $O_4$ one should be able to
extract the values of the coefficents $A_{nm}$ and, where possible, perform
exponentiation to resum the infrared logarithms.

\section*{Acknowledgements}

We thank Walter Giele for useful and stimulating discussions.  We also thank
Siggi Bethke for pointing out that the JADE-E0 jet algorithm does not use the
E0 recombination scheme. JMC and MC thank the UK Particle Physics and
Astronomy  Research Council for research studentships. EWNG thanks the theory
groups at CERN and Fermilab for their kind hospitality during the early stages
of this work. This work was supported in part by the EU Fourth Framework 
Programme `Training and Mobility of Researchers',  Network `Quantum
Chromodynamics and the Deep Structure of  Elementary Particles', contract
FMRX-CT98-0194 (DG-12 - MIHT).

\newpage
\appendix
\section{Cancellation of infrared singularities}
\setcounter{equation}{0}
\label{sec:app}
In order to compute next-to-leading order quantities in perturbation
theory, it is necessary to combine the contribution from $n$-parton
one-loop Feynman diagrams with the $(n+1)$-parton bremstrahlung process.
The virtual matrix elements are divergent and contain both infrared
and ultraviolet singularities.  The ultraviolet poles are removed by
renormalisation, however the soft and collinear infrared poles are
only cancelled when the virtual graphs are combined with the
bremstrahlung process.  Although the cancellation of infrared poles
can be done analytically for simple processes, for complicated
processes, it is necessary to resort to numerical
techniques.  

\subsection{A simple example}
\label{subsec:simplex}

The numerical problem has been nicely formulated by Kunszt and Soper
\cite{KS} by means of a simple example integral,
\begin{equation}
{\cal I} = \lim_{\epsilon \to 0} \left \lbrace
\int^1_0 \frac{dx}{x} x^\epsilon F(x) -
\frac{1}{\epsilon} F(0)\right \rbrace,
\end{equation}
where $F(x)$ is a known but complicated function representing the
$(n+1)$-parton bremstrahlung matrix elements.  Here $x$ represents the
angle between two partons or the energy of a gluon and the integral
over $x$ represents the additional phase space of the extra parton.
As $x\to 0$, the integrand is regularised by the $x^\epsilon$ factor
as in dimensional regularisation, however, the first term is still
divergent as $\epsilon \to 0$.  This divergence is cancelled by the
second term - the $n$-parton one-loop contribution - so that the
integral is finite.  A variety of methods to compute ${\cal I}$ have
been developed.

The method used by Ellis, Ross and Terrano \cite{ert} is also known as
the subtraction method.  Here, a divergent term is subtracted from the
first term and added to the second, 
\begin{eqnarray} 
{\cal I} &= & \lim_{\epsilon \to 0}
\left \lbrace \int^1_0 \frac{dx}{x} x^\epsilon (F(x)-F(0)) 
+F(0) \int^1_0 \frac{dx}{x} x^\epsilon 
- \frac{1}{\epsilon} F(0)\right \rbrace \nonumber \\ 
&=& \int^1_0 \frac{dx}{x}  \left (F(x)-F(0)\right), 
\end{eqnarray} 
so that the integral is manifestly finite.  This method has the
advantages of requiring (in principle) no extra theoretical cutoffs and
making no approximations.   However, in practice, there are still large
cancellations in the numerator and there is a hidden parameter which
cuts the integral off at the lower end. Recently, this technique has
been developed    to describe general processes \cite{subtract}.

An alternate approach known as the phase space slicing method has also
been widely used \cite{slice}.  The integration region is divided into
two parts, $0 < x < \delta$ and $\delta < x < 1$.  In the first region,
the function $F(x)$ can be approximated by $F(0)$ provided the
arbitrary cutoff $\delta \ll 1$,
\begin{eqnarray} 
{\cal I} &\sim & \lim_{\epsilon \to 0} \left \lbrace
\int^1_\delta \frac{dx}{x} x^\epsilon F(x) +F(0) \int^\delta_0 \frac{dx}{x}
x^\epsilon - \frac{1}{\epsilon} F(0)\right \rbrace \nonumber \\ &\sim &
\int^1_\delta \frac{dx}{x}   F(x) +F(0) \ln(\delta). \label{eq:slice}
\end{eqnarray} 
This method is extremely portable \cite{GG} since the soft and
collinear approximations of the matrix elements and phase space are
universal.  This makes it easy to apply to a wide variety of physically
interesting processes.  However, the disadvantage is the presence of
the arbitrary cutoff $\delta$.  The integral should not depend on
$\delta$, and the $\delta$ dependence of the two terms in
eq.~(\ref{eq:slice}) should cancel.  Since the approximations are
reliable only when $\delta$ is small, this can give rise to numerical
problems.

Finally, we can imagine combinations of these two approaches - the
hybrid approach.  There are two scales in the problem, $\delta$ and
$\Delta$. In the region $0 < x < \delta$, we adopt the slicing
procedure, while in the range $\delta < x < \Delta$ we add and subtract
an analytically integrable set of universal terms,  $E(x)$, to
eq.~(\ref{eq:slice}),
\begin{equation} 
{\cal I} \sim    \int^1_\delta \frac{dx}{x}   F(x) +F(0)
\ln(\delta) - \int^\Delta_\delta \frac{dx}{x} E(x)   + \int^\Delta_\delta
\frac{dx}{x} E(x),
\end{equation}
which on rearrangement yields,
\begin{equation}
{\cal I} \sim  
\int^1_\delta \frac{dx}{x}  \left(
F(x)-E(x)\Theta(\Delta-x)  \right) 
+\int^\Delta_\delta \frac{dx}{x}  E(x) +F(0) \ln(\delta).
\label{eq:hybrid}
\end{equation} 
Because we explicitly add and subtract the same quantity, there can be
no dependence on $\Delta$ which can therefore be taken to be large.
However, the slicing approximation requires $\delta \to 0$. For this
approach to be useful, two conditions must be satisfied. First, the
second term in  eq.~(\ref{eq:hybrid})  must be evaluated analytically 
without making any approximation in the phase space and should produce
a term $-F(0) \ln(\delta)$ from the lower boundary that explicitly
cancels the third (slicing) term. This allows the limit $\delta \to 0$ 
to be taken (inasmuch as that can be achieved numerically). Second,
$F(x) \sim E(x)$ as $x \to 0$ and more usefully $E(x)$ is smooth and as
close to $F(x)$ as possible over the whole range of  $x < \Delta$, so
that the first term in eq.~(\ref{eq:hybrid}) can be safely evaluated
numerically. This is the technique we have adopted in this paper. 

\subsection{Singular behaviour of Matrix Elements}
\label{subsec:appant}

Clearly the choice of the subtraction function $E(x)$ requires some
care, as does the integration over the phase space variables $x$.   To
help us do this in a sensible way, we first  recall the well known
singular behaviour of the matrix elements. This is most clearly seen by
decomposing the amplitude according  to the various colour structures.
For example,  in the process, $$e^+e^- \to q\bar q + n~g,$$ the
amplitude can be written as, 
\begin{equation}
\M(Q_1,\Qbar_2;1,\ldots,n) =
\widehat{\S}^{n+2}_\mu(Q_1;1,\ldots,n;\overline{Q}_2)V^\mu,
\end{equation} 
where the hadronic current is given by, 
\begin{equation}
\widehat{\S}^{n+2}_\mu(Q_1;1,\ldots,n;\overline{Q}_2)
=ieg^n\sum_{P(1,\ldots,n)} (T^{a_1}\ldots T^{a_n})_{c_1c_2}
S_\mu(Q_1;1,\ldots,n;\overline{Q}_2). 
\label{eq:ngcur} \end{equation}
Here, $\S_\mu(Q_1;1,\ldots,n;\overline{Q}_2)$ represents the colourless
subamplitude where the gluons (with colours $a_1,\ldots,a_n$) are
emitted in an ordered way from the quarks (with colours $c_1$ and
$c_2$). The colour matrices are normalised such that, $$ {\rm Tr}
\left(T^{a_i}T^{a_j}\right) = \frac{1}{2}\delta^{a_ia_j}. $$ By summing
over all permutations of gluon emission, all Feynman diagrams and
colour structures are accounted for. On squaring, we find that for $n
\geq 1$ the leading colour piece is simply, 
\begin{equation} 
\Big
|\widehat{\S}^{n+2}_\mu V^\mu\Big |^2 = e^2
\left(\frac{g^2N}{2}\right)^n \left(\frac{N^2-1}{N}\right)
\sum_{P(1,\ldots,n)}  \left( \Big
|\S_\mu(Q_1;1,\ldots,n;\overline{Q}_2) V^\mu\Big |^2  + {\cal O}\left(
\frac{1}{N^2}\right)\right). 
\end{equation} 
The subleading terms are slightly more complicated, but may be
straightforwardly obtained by considering diagrams with gluons replaced
by one or more  photons.\footnote{Here, we have focussed on the two
quark process, however, the same  type of colour decomposition can be
applied to the four quark process,  $e^+e^- \to q\bar q Q\Qbar +
(n-2)g,$ (see for example Ref.~\cite{triple,JMC}). The structure
appears to be more complicated, however the singular behaviour  of
individual contributions follows the same pattern.}

The advantage of the colour decomposition is that the ordered
subamplitudes have particularly simple singular limits. For example, in
the limit where gluon $u$ is soft, we have the QED-like factorisation
into an eikonal factor multiplied by the colour ordered amplitude with
gluon $u$ removed, but the ordering of the hard gluons preserved,
\begin{eqnarray}
\lefteqn{
\Big |\S_\mu(Q_1;1,\ldots,a,u,b,\ldots,n;\overline{Q}_2) V^\mu\Big |^2 }
\\
&\to& S_{aub}(s_{ab},s_{au},s_{ub})\,
\Big |\S_\mu(Q_1;1,\ldots,a,b,\ldots,n;\overline{Q}_2) V^\mu\Big
|^2,
\nonumber
\end{eqnarray}
with the eikonal factor given by,
\begin{equation}
S_{aub}(s_{ab},s_{au},s_{ub})=\frac{4s_{ab}}{s_{au}s_{ub}}.
\label{eq:ssoft}
\end{equation}
Similarly, in the limit where two particles become collinear, the
sub-amplitudes factorise. For example, if gluons $a$ and $b$ become
collinear and form gluon $c$, then only colour connected gluons give a
singular contribution. For example, 
\begin{equation}
\Big |\S_\mu(Q_1;1,\ldots,a,b,\ldots,n;\overline{Q}_2) V^\mu\Big |^2 
\to P_{gg\to g}(z,s_{ab})\,
\Big |\S_\mu(Q_1;1,\ldots,c,\ldots,n;\overline{Q}_2) V^\mu\Big |^2.
\end{equation}
For gluons that are not colour connected,  there is no singular
contribution as $s_{ab} \to 0$, and, when integrated over the small
region of phase space where the collinear approximation is valid, give
a negligible contribution to the cross section. Here $z$ is the
fraction of the momentum carried by one of the gluons and, after
integrating over the azimuthal angle of the plane containing the
collinear particles with respect to the rest of the hard process, the
collinear splitting function $P_{gg\to g}$ is given by,
\begin{equation}
P_{gg\to g} (z,s) = \frac{2}{s} P_{gg\to g} (z)
\label{eq:double}
\end{equation}
where the usual Altarelli-Parisi splitting kernel
with the colour factor removed is given by \cite{AP},
\begin{equation}
P_{gg\to g} (z) = \left( \frac{1+z^4+(1-z)^4}{z(1-z)}\right).
\label{eq:pgg}
\end{equation}
Similar splitting kernels exist  
for other combinations of collinear partons \cite{AP},
\begin{eqnarray}
P_{qg \rightarrow q}(z)&=&
 \left( \frac{1+z^2}{1-z} \right) , \\
P_{q\bar{q} \rightarrow g}(z)&=&
 \left(z^2+(1-z)^2\right),
\label{eq:pqg}
\end{eqnarray}
with $P_{gq \to q}(z) = P_{qg \to q}(1-z)$. As before, the colour
factors have been removed and azimuthal averaging of the collinear
particle plane is understood.

In both the soft and collinear limits, the colour ordered squared
amplitudes factorise into a squared amplitude containing one less
parton multiplied by a factor that depends on the the unresolved
particle and the two adjacent `hard' particles.   We view the two
`hard' particles as an {\em antenna} from which the unresolved parton
is radiated.  It therefore makes sense to divide the phase space in a
similar way and to treat the subtraction term as the singular factor
for the whole antenna integrated over the unresolved phase space.

\subsection{Phase space factorisation}
\label{subsec:appps}

Let us consider an $(n+1)$ particle phase space described by momenta
$p_i$ with $p_i^2 = 0$ for $i=1,\ldots,n$ . If the total centre of mass
energy is $Q$, then let us denote the phase space  by, $\d
PS(Q^2;p_1,\ldots,p_n)$. As discussed above, we wish to relate the full
$(n+1)$ particle phase space to an $n$ particle phase space whenever
one of the original $(n+1)$  particles is unresolved. Let the
unresolved particle be labelled by $u$ and the two adjacent hard
particles by $a$ and $b$, then the phase space can be factorised as,
\begin{equation}
\d PS(Q^2;p_1,\ldots,p_n) = \d PS(Q^2;p_1,\ldots,p_{aub},\ldots,p_n)
~\frac{\d s_{aub}}{2\pi}
~\d PS(s_{aub};p_a,p_u,p_b),
\label{eq:ps}
\end{equation}
where $p_{aub} = p_a+p_u+p_b$ and $p_{aub}^2 = s_{aub}$. To factorise
the phase space into an $n$ particle phase space  multiplied by a
factor containing integrals over the unresolved invariants $s_{au}$ and
$s_{ub}$ that appear in the singular limits of the matrix elements, we
multiply the r.h.s. of eq.~(\ref{eq:ps}) by,
\begin{equation}
\d PS(s_{AB};p_A,p_B) ~/ \int \d PS(s_{AB};p_A,p_B),
\end{equation}
where particles $A$ and $B$ have momenta $p_A$ and $p_B$ such that,
$p_{aub}= p_{AB} = p_A+p_B$, $p_A^2 = p_B^2 = 0$ and $s_{aub} =
s_{AB}$. In other words,
\begin{eqnarray}
\d PS(Q^2;p_1,\ldots,p_n) &=& \d PS(Q^2;p_1,\ldots,p_{AB},\ldots,p_n)
~\frac{\d s_{AB}}{2\pi}
~\d PS(s_{AB};p_A,p_B) \times \d PS^{\rm sing}  \nonumber \\
&=& \d PS(Q^2;p_1,\ldots,p_A,p_B,\ldots,p_n) \times \d PS^{\rm sing}.
\label{eq:psfac}
\end{eqnarray}
As desired, we have the phase space for an final state  containing $n$
lightlike particles multiplied  by $\d PS^{\rm sing}$. Working in
four-dimensions and after integration over the Euler angles, 
\begin{eqnarray}
\d PS^{\rm sing} &=& 
\frac{\d PS(s_{aub};p_a,p_u,p_b)}{ \int \d PS(s_{AB};p_A,p_B)}\nonumber \\
&=&
\frac{1}{16\pi^2}s_{aub}
\d x_{au}\d x_{ub}\d x_{ab}\delta(1-x_{au}-x_{ub}-x_{ab}),
\end{eqnarray}
where $x_{ij} = s_{ij}/s_{aub}$. For this to work, a mapping must exist
that determines $p_A$ and $p_B$ for  a given set of momenta $p_a$,
$p_b$ and $p_u$. Many choices are possible \cite{event2,kosower} and we
choose the symmetric mapping of \cite{kosower},
\begin{eqnarray}
p_{A} &=& 
\frac{1}{2}\left[1+\rho+\frac{s_{ub}(1+\rho-2r_{1})}{s_{ab}+s_{au}}\right]p_{a}
+r_{1}p_{u}
+\frac{1}{2}\left[1-\rho+\frac{s_{au}(1-\rho-2r_{1})}{s_{ab}+s_{ub}}\right]p_{b},
\nonumber \\
p_{B} &=& 
\frac{1}{2}\left[1-\rho-\frac{s_{ub}(1+\rho-2r_{1})}{s_{ab}+s_{au}}\right]p_{a}
+(1-r_{1})p_{u} 
+\frac{1}{2}\left[1+\rho-\frac{s_{au}(1-\rho-2r_{1})}{s_{ab}+s_{ub}}\right]p_{b},
\nonumber \\
\label{eq:transform}
\end{eqnarray}
where, 
\begin{equation}
r_{1}=\frac{s_{ub}}{s_{au}+s_{ub}},
\end{equation}
and,
\begin{equation}
\rho=\sqrt{\frac{s_{ab}^{2}+(s_{au}+s_{ub})s_{ab}+4r_{1}(1-r_{1})s_{au}s_{ub}}
{s_{ab}s_{aub}}}.
\end{equation}
Note that this transformation approaches the singular limits smoothly.
For example, as $s_{au} \to 0$, then $r_1 \to 1$, $\rho \to 1$ and $p_A
\to p_a+p_u$, $p_B \to p_b$. 

\subsection{Antenna factorisation of the Matrix elements}
\label{sec:appsub}

Having factorised the phase space, we now wish to find the analogues of
the subtraction functions $E(x)$ discussed in
Appendix~\ref{subsec:simplex}.    These functions should ideally be
valid over the whole of the antenna phase space $\d PS^{\rm sing}$ and,
in the soft and collinear regions must match onto the singular limits
discussed in Appendix~\ref{subsec:appant}. In other words, for a given
$(n+1)$ particle amplitude,  in the limit where particle $u$ is
unresolved,
\begin{equation}
\Big |\S_\mu(\ldots,a,u,b,\ldots) V^\mu\Big |^2 \to 
\A_{aub} \,
\Big |\S_\mu(\ldots,A,B,\ldots) V^\mu\Big |^2,
\label{eq:Adef}
\end{equation}
where we have replaced the antenna comprising $a,u,b$ by the hard
partons $A$ and $B$  to obtain an $n$ particle amplitude. The antenna
function $\A_{aub}$ depends on the momenta of the radiated particles
$a$, $b$ and $u$, but the $n$ particle amplitude $|\S_\mu V^\mu |^2$
does not.

The leading colour contribution to an observable cross section from an
$(n+1)$ particle final state with a particular colour ordering is
proportional to, 
\begin{equation}
\left(\frac{N^2-1}{N^2}\right)
\g2No2^{n+1}\Big |\S_\mu(\ldots,a,u,b,\ldots) V^\mu\Big |^2~\J_{(n+1)}   
~ \d PS (Q^2;\ldots,p_a,p_u,p_b,\ldots), 
\end{equation} 
where the  observable function $\J_{(n+1)}$ represents the  cuts
applied to the $(n+1)$ particle  phase space to define the observable.
Using the factorisation of the matrix elements defined in
eq.~(\ref{eq:Adef}),  when particle $u$ is unresolved we should
subtract, 
\begin{eqnarray} 
\g2No2^{n+1} \A_{aub} ~\Big
|\S_\mu(\ldots,A,B,\ldots) V^\mu\Big |^2 ~\J_{(n)}  ~\d PS
(Q^2;\ldots,p_a,p_u,p_b,\ldots), 
\label{eq:subtract} 
\end{eqnarray} 
from the $(n+1)$ particle contribution and, using the phase space
factorisation of eq.~(\ref{eq:psfac}), add, 
\begin{eqnarray}
\g2No2^{n+1} \A_{aub}~\d PS^{\rm sing} ~\Big |\S_\mu(\ldots,A,B,\ldots)
V^\mu\Big |^2 ~\J_{(n)}  ~\d PS (Q^2;\ldots,p_A,p_B,\ldots), 
\label{eq:add}
\end{eqnarray} 
to the $n$ particle contribution where both the observable function 
$\J$ and matrix elements $|\S_\mu V^\mu |^2$  depend only on the
momenta of the  $n$ remaining hard partons.  Note that for any infrared
safe observable, in the limit that one particle is unresolved,
$\J_{(n+1)} \to \J_{(n)}$.  In the subtraction term
eq.~(\ref{eq:subtract}), we use the transformations of 
eq.~(\ref{eq:transform}) to map the momenta $p_a$, $p_u$ and $p_b$
defined  in the $(n+1)$ particle phase space onto the momenta $p_A$ and
$p_B$ used in the $n$-particle  matrix elements and observable
functions. In eq.~(\ref{eq:add}), all dependence on the momenta of 
particles $a$, $b$ and $u$ may be integrated out to give the antenna
factor, $\F$, 
\begin{equation} 
\F_{AB}(s_{AB}) = \g2No2 \int \A_{aub}~\d PS^{\rm sing},
\end{equation} 
multiplying  the $n$ particle cross section (for a given colour ordered
amplitude), 
\begin{eqnarray} 
\g2No2^{n} \Big |\S_\mu(\ldots,A,B,\ldots)
V^\mu\Big |^2 ~\J_{(n)}  ~\d PS (Q^2;\ldots,p_A,p_B,\ldots). 
\end{eqnarray} 
The full set of subtraction terms is obtained by summing over all
possible antennae. 

The Dalitz plot for the $(AB) \to aub$ phase space is shown in 
Fig.~\ref{fig:dalitz}.   In the hybrid scheme we are implementing,  we
use the slicing method of \cite{GG} in the region  $\min(s_{au},s_{ub})
< \delta$, and the subtraction scheme in the region, $\delta <
\min(s_{au},s_{ub}) < \Delta$. In the slicing region, the phase space
and soft and collinear approximations  to the matrix elements are kept
in $D=4-2\e$  dimensions to regularise the singularities present when
either invariant vanishes.   Using the approach of \cite{GG}, there are
three separate contributions (a) soft gluon when $\max(s_{au},s_{ub}) <
\delta$, (b) $a$ and $u$ collinear when $s_{au} < \delta$ but $s_{ub} >
\delta$ and (c) $b$ and $u$ collinear when $s_{ub} < \delta$ but
$s_{au} > \delta$.

Before turning to the explicit forms for the antenna subtraction terms,
we note that while quarks are only directly colour connected to one
particle  - a gluon or antiquark, gluons are directly connected to two
particles  - the gluon (or quark) on either side. Therefore,while the
quark (or antiquark) appear in a single antenna, gluons appear in two. 
This gives an ambiguity in how to assign the collinear  singularities
of a pair of gluons to each antenna.   Later we will exploit this 
ambiguity to make the antenna functions $\A_{aub}$  for different pairs
of hard partons finite simpler.  

\begin{figure}[t]
\begin{center}
~\psfig{file=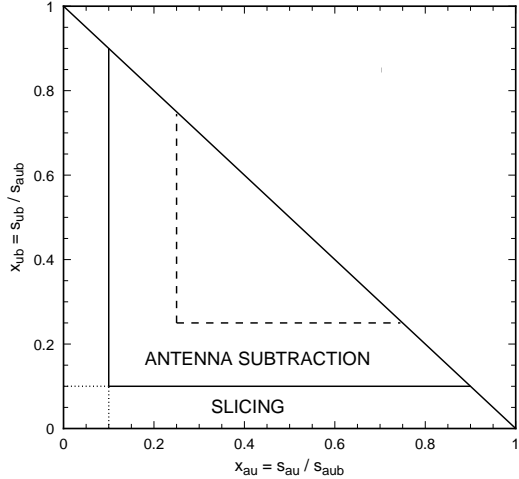, width=7cm,angle=90} 
\caption{The phase space for the decay $(AB) \to aub$.   The cut 
$\min(s_{au},s_{ub}) = \delta$ with $\delta = 0.1~s_{aub}$ is shown
as a solid line while $\min(s_{au},s_{ub}) = \Delta$ is shown as a
 dashed line for $\Delta = 0.25~s_{aub}$.   
 The region $\min(s_{au},s_{ub}) < \delta$ 
defines where the slicing approach is utilised, with the soft and 
collinear regions demarked by dotted lines.   Antenna subtraction is applied 
when $\delta < \min(s_{au},s_{ub}) < \Delta$.}
\label{fig:dalitz}
\end{center}
\end{figure}  

\subsubsection{Quark-Antiquark antenna}  \label{subsubsec:QQbar}

Let us first consider a system containing a quark, antiquark  and a
gluon. This is produced by an antenna comprising of a hard quark and
antiquark pair that decays by radiating a gluon.  Any function that has
the correct soft gluon and collinear quark/gluon singularities in the
appropriate limit is satisfactory. Here the hard particles in the
antenna are $Q$ and $\overline{Q}$ which radiate to form $q$, $\bar q$
and the gluon $g$.  A suitable choice for the antenna function is, 
\begin{eqnarray}  
\A_{qg\bar q} &=&   \frac{|\S_\mu(q;g;\bar q)
V^\mu|^2} {|\S_\mu(Q;\Qbar) V^\mu|^2} \nonumber \\ &=& 
\frac{2}{s_{aub}} \left(\frac{x_{au}}{x_{ub}} +\frac{x_{ub}}{x_{au}}
+\frac{2x_{ab}x_{aub}}{x_{au}x_{ub}} \right). 
\end{eqnarray} 
Because this is proportional to the three parton matrix elements,
$|\S_\mu(q;g;\bar q) V^\mu|^2$,  it automatically contains the correct
soft and collinear limits. Furthermore, it is smooth over the whole
three particle phase space and singularities only appear in the $s_{au}
\to 0$ and $s_{ub}\to 0$ limits.

Explicitly integrating over the antenna phase space for 
$\delta < \min(s_{au},s_{ub}) < \Delta$  we find,
\begin{eqnarray}
\F_{Q\Qbar}(s_{Q\Qbar}) 
&=& \g2No2 \int \A_{qg\bar q}~\d PS^{\rm sing}\nonumber \\
&=& 
\left(\frac{\alpha_sN}{2\pi}\right)
\Biggl(
\ln^2\left(\frac{\delta}{s_{Q\Qbar}}\right)
+\frac{3}{2}\ln\left(\frac{\delta}{s_{Q\Qbar}}\right)  \Biggr )
+\F_{Q\Qbar}^{\Delta}\left(\frac{\Delta}{s_{Q\Qbar}}\right)
+ {\cal O}(\delta)  .
\end{eqnarray}
Since we intend to take the $\delta \to 0$ limit, the terms of ${\cal
O}(\delta)$ may be safely neglected. The $\delta$ independent function
$\F_{Q\Qbar}^{\Delta}$ is given by,
\begin{equation}
\F_{Q\Qbar}^{\Delta}(x) 
=
\left(\frac{\alpha_sN}{2\pi}\right)
\Biggl(
-\ln^2\left(x\right)  
+ \frac{5x}{2} -2 \Li_2(x)
+\left(\frac{3}{2}-2x+\frac{x^2}{2}\right)
\ln\left(\frac{1-x}{x}\right) \Biggr ).
\label{eq:FQQbarDelta}
\end{equation}

\subsubsection{Quark-Gluon antenna} 
\label{subsubsec:QG}

For antenna made of a quark $Q$ and gluon $G$, there are two possible
ways of radiating.   Either a gluon can be radiated so that a
quark-gluon-gluon system is formed, or the gluon may split into a
antiquark-quark pair. This latter possibility is subleading in the
number of colours and the  discussion of situations like this is
deferred to sec.~\ref{subsubsec:qqbar}.

For a quark-gluon-gluon system there is a less obvious choice of
antenna function, particularly since the singularity that is produced
when the gluon splits sits in more than one antenna.   If, in the
collinear limit, the gluon splits into an unresolved gluon $u$ which
carries momentum fraction $z$ and a hard gluon $b$ with momentum
fraction $1-z$, the antenna function  should naively be proportional to
$P_{gg\to g}$ which is singular as $z \to 0$  and $z \to 1$.  This
corresponds to singularities  as both $s_{ub} \to 0$ and $s_{ab} \to
0$. However, because the collinear singularity sits in more than one
antenna -  the two gluons also occur in a second antenna where the role
of the two gluons is reversed - we can make use of the $N=1$
supersymmetric identity to rewrite $P_{gg\to g}$ as,
\begin{equation} 
P_{gg\to g} = P_{qg\to q} +
P_{gq\to q }-P_{q\bar q\to g}. 
\end{equation} 
The soft singularities as $z \to 0$ are contained in $P_{gq\to q}$
while those  as $z \to 1$ are in $P_{qg\to q}$.  We therefore divide
$P_{gg\to g}$ amongst the two antennae such that $P_{gq\to q}$ sits  in
the antenna where gluon $u$ is unresolved. The $z \to 1$ singularities
are placed in the antenna where the role of the two gluons is
reversed.   The remaining $P_{q\bar q\to g}$ may be divided between the
two antennae according to choice.  With a slight modification due to
the $P_{q\bar q\to g}$ term, the antenna  function used for the
$Q\Qbar$ antenna has the correct limits, so that,
\begin{equation} 
\A_{qgg} 
=
\A_{qg\bar q}
-
\frac{2}{s_{aub}}
\left(\frac{x_{au}^2}{x_{ub}x_{aub}}\right).
\end{equation}
This is again smooth over the whole three particle phase space with
singularities only appearing  in the $s_{au} \to 0$ and $s_{ub} \to 0$
limits. In particular, as $z \to 0$, the collinear limit matches onto
the soft limit which would not have been the case if we had divided the
soft/collinear singularities equally between the two antenna.

After integrating over the antenna phase space for  $\delta <
\min(s_{au},s_{ub}) < \Delta$  we find,
\begin{eqnarray}
\F_{QG}(s_{QG}) &=& \g2No2 \int \A_{qgg}~\d PS^{\rm sing}\nonumber \\ 
&=&
\left(\frac{\alpha_sN}{2\pi}\right)
\Biggl(
\ln^2\left(\frac{\delta}{s_{QG}}\right)
+\frac{10}{6}\ln\left(\frac{\delta}{s_{QG}}\right)  \Biggr )
+\F_{QG}^{\Delta}\left(\frac{\Delta}{s_{QG}}\right)
\end{eqnarray}
with the
$\delta$ independent function $\F_{QG}^{\Delta}$ is given by,
\begin{eqnarray}
\F_{QG}^{\Delta}(x) 
&=&
\left(\frac{\alpha_sN}{2\pi}\right)
\Biggl(
-\ln^2\left(x\right)  
+ \frac{19x}{6} 
-\frac{x^2}{6}
+\frac{x^3}{9} 
 -2 \Li_2(x)\nonumber \\
&&~~~~~~~~~~~
+\left(\frac{10}{6}-2x+\frac{x^2}{2}
-\frac{x^3}{6}\right)
\ln\left(\frac{1-x}{x}\right) \Biggr ).
\label{eq:FQGDelta}
\end{eqnarray}

Antennae containing a gluon and an antiquark are described by,
\begin{equation}
\A_{gg\bar q} = \A_{qgg}( a \leftrightarrow b),
\end{equation}
and,
\begin{equation}
\F_{G\Qbar}(s_{G\Qbar}) = \F_{QG}(s_{G\Qbar}).
\end{equation}

\subsubsection{Gluon-Gluon antenna} 
\label{subsubsec:GG}
For antenna comprising only gluons, we repeat this SUSY inspired
trick for each 
of the resolved gluons so that,
\begin{equation} 
\A_{ggg} 
=
\A_{qg\bar q}
-\frac{2}{s_{aub}} 
\left(\frac{x_{au}^2}{x_{ub}x_{aub}}
+\frac{x_{ub}^2}{x_{au}x_{aub}}\right).
\end{equation}
Note that Kosower~\cite{kosower} has proposed an antenna factorisation
for gluonic processes,
\begin{equation}
\A_{ggg}^{\rm Kosower} 
= \frac{4}{s_{aub}} 
\left(\frac{(x_{aub}(x_{aub}-x_{ab})+x_{ab}^2)^2}
{x_{au}x_{ub}x_{ab}x_{sub}}\right),
\end{equation}
which, in the $u/b$ collinear limit  regenerates the full $P_{gg \to g}$
splitting function,  as well as the soft limits.

Integration of the antenna function $\A_{ggg}$
over the whole of the subtraction region yields,
\begin{eqnarray}
\F_{GG}(s_{GG}) &=& \g2No2 \int \A_{ggg}~\d PS^{\rm sing}\nonumber \\ 
&=&
\left(\frac{\alpha_sN}{2\pi}\right)
\Biggl(
\ln^2\left(\frac{\delta}{s_{GG}}\right)
+\frac{11}{6}\ln\left(\frac{\delta}{s_{GG}}\right)  \Biggr )
+\F_{GG}^{\Delta}\left(\frac{\Delta}{s_{GG}}\right)
\end{eqnarray}
with the
$\delta$ independent function $\F_{GG}^{\Delta}$ is given by,
\begin{eqnarray}
\F_{GG}^{\Delta}(x) 
&=&
\left(\frac{\alpha_sN}{2\pi}\right)
\Biggl(
-\ln^2\left(x\right)  
+ \frac{23x}{6} 
-\frac{2x^2}{6}
+\frac{2x^3}{9} 
 -2 \Li_2(x)\nonumber \\
&&~~~~~~~~~~~
+\left(\frac{11}{6}-2x+\frac{x^2}{2}
-\frac{x^3}{3}\right)
\ln\left(\frac{1-x}{x}\right) \Biggr ).
\label{eq:FGGDelta}
\end{eqnarray}

\subsubsection{Antenna where a quark-antiquark pair merge} 
\label{subsubsec:qqbar}

There are also configurations when two (or more) colour lines are
present,  one ending in an antiquark the other starting  with a quark
of the same flavour. Here the matrix elements have the form, 
\begin{equation} 
\Big | \S_\mu(\ldots,a,\bar q | q,b,\ldots)V^\mu\Big
|^2.  
\end{equation}  
In the collinear limit, the quark-antiquark pinch
the two colour lines together to form a single colour line, 
\begin{equation} 
\Big | \S_\mu(\ldots,a,\bar q| q,b,\ldots)V^\mu\Big
|^2  \to  P_{\bar q q \to g}(z,s_{\bar q q})  \Big |
\S_\mu(\ldots,a,G,b,\ldots)V^\mu\Big |^2, 
\end{equation} 
with $P_{\bar q q \to G}(z,s)$ given by eqs.~(\ref{eq:double})  and
(\ref{eq:pqg}). There is no soft singularity, nor is there any
dependence on the type  of adjacent parton, $a$ or $b$. Clearly, the
quark-antiquark pair can sit in two antennae, $(a,\bar q,q)$ and $(\bar
q, q,b)$ and we have some freedom of how  to assign the singularities
to the antennae. There are two obvious choices.   Either we divide the
singular contribution  equally  over the two antennae, or, we place the
$z^2$ part of $P_{\bar q q \to g}(z)$ in one antenna and the $(1-z)^2$
part in the other (as we did with the  three gluon antenna before). 
While there appears to be no preference,  we follow this latter route
so that the antenna function vanishes as the unresolved particle
becomes soft,
\begin{equation}  
\A_{a\bar q q}  = \frac{2}{s_{a\bar q q}} 
\left(\frac{x_{a\bar q}^2}{x_{\bar q q}x_{a\bar q q}}\right),
\end{equation} 
and, 
\begin{equation} 
\A_{\bar q q b}  = \A_{a\bar q q}
(x_{a\bar q}  \to x_{qb}, x_{a\bar q q} \leftrightarrow x_{\bar q q
b}). 
\end{equation}

Following this procedure and integrating over the whole of the
subtraction region yields,
\begin{eqnarray}
\F^{N_F}_{aG}(s_{aG}) &=& 
\left(\frac{g^2N_F}{2}\right)
\int \A_{a\bar q q}~\d PS^{\rm sing}\nonumber \\ 
&=&
\left(\frac{\alpha_sN_F}{2\pi}\right)
\Biggl(
-\frac{1}{6}\ln\left(\frac{\delta}{s_{aG}}\right)  \Biggr )
+\F_{aG}^{N_F\Delta}\left(\frac{\Delta}{s_{aG}}\right),
\end{eqnarray}
and,
\begin{equation}
\F^{N_F}_{Gb}(s_{Gb}) = \F^{N_F}_{aG}(s_{Gb}).
\end{equation}
The factor of $N_F$ arises because 
each of the $N_F$ quark flavours may contribute.
The $\delta$ independent function is,
\begin{eqnarray}
\F_{aG}^{N_F\Delta}(x) &=& \F_{Gb}^{N_F\Delta}(x) \nonumber \\
&=&
\left(\frac{\alpha_sN_F}{2\pi}\right)
\Biggl(
- \frac{2x}{3} + \frac{x^2}{6} - \frac{x^3}{9} 
-\left(\frac{1}{6}-\frac{x^3}{6}\right)
\ln\left(\frac{1-x}{x}\right) \Biggr ).
\label{eq:FqqbarDelta}
\end{eqnarray}

\subsection{Leading colour contribution to $e^+e^- \to 4$~jets.} 

As a pedagogical example, we consider the leading colour contribution
relevant for $e^+e^- \to 4$~jets. The sub-leading pieces are similarly
calculated but the resulting expressions are somewhat lengthy due to
the many antennae that are involved~\cite{JMC}. At leading order in the
number of colours, only the two quark and $n$ gluon process
contributes\footnote{The four quark process gives a contribution that
is suppressed by a factor of $N_F/N$ relative to the leading colour
contribution.  Numerically this is an important contribution and,
together with the other subleading terms, is included in the numerical
results presented earlier.}, so, at lowest order, the cross section is
given by, 
\begin{eqnarray} 
\frac{\d \sigma_{4}^{\rm LO}}{\sigma_0} 
&=&\frac{(2\pi)^5}{s} \left(\frac{N^2-1}{N^2}\right) \asN^2 \nonumber
\\ 
&\times &\sum_{P(G_1,G_2)} \Big |\S_\mu(Q_1;G_1,G_2;\Qbar_2)
V^\mu\Big |^2 ~\J_{(4)}  ~\d PS (Q^2;Q_1G_1,G_2,\Qbar_2) \I_2,
\label{eq:LO} \end{eqnarray} 
where $\I_2$ is the identical particle factor for the two gluon final
state. In practice, the $2!$ permutations precisely cancels the
identical particle factor of $1/2!$, and it is more convenient to keep
one particular ordering so that,
\begin{equation} 
\frac{\d \sigma_{4}^{\rm LO}}{\sigma_0} =
\frac{(2\pi)^5}{s} \left(\frac{N^2-1}{N^2}\right) \asN^2 \Big
|\S_\mu(Q_1;G_1,G_2;\Qbar_2) V^\mu\Big |^2 ~\J_{(4)}  ~\d PS
(Q^2;Q_1G_1,G_2,\Qbar_2). 
\end{equation}

Similarly, the leading colour contribution from the five parton
bremstrahlung process is, 
\begin{eqnarray}  
\frac{\d\sigma_{5}}{\sigma_0} &=& \frac{(2\pi)^7}{s}
\left(\frac{N^2-1}{N^2}\right) \asN^3 \nonumber \\ 
&\times &
\sum_{P(g_1,g_2,g_3)}  
\Big |\S_\mu(q_1;g_1,g_2,g_3;\bar q_2) V^\mu\Big|^2 ~\J_{(5)}  
~\d PS (Q^2;q_1,g_1,g_2,g_3,\bar q_2) \I_3,\nonumber \\
&=& \frac{(2\pi)^7}{s} \left(\frac{N^2-1}{N^2}\right) \asN^3 
\Big |\S_\mu(q_1;g_1,g_2,g_3;\bar q_2) V^\mu\Big |^2 ~\J_{(5)}  
~\d PS (Q^2;q_1,g_1,g_2,g_3,\bar q_2).	
\nonumber \\ \label{eq:sig5} 
\end{eqnarray}  
Note that $\J_5$ projects the five parton momenta onto the four jet
like observable. Once again, we can cancel the identical particle
factor $\I_3 = 1/3!$ against the $3!$ permutations of the gluons, and
retain the single permutation indicated. For this colour ordering,
three antennae will contribute, $(q_1,g_1,g_2)$,  $(g_1,g_2,g_3)$ and
$(g_2,g_3,\bar q_2)$ where in each case the parton in the middle is
unresolved.  In the first antenna, $\{p_{q_1},p_{g_1},p_{g_2}\} \to
\{p_{Q_1},p_{G_1}\}$ according to eq.~(\ref{eq:transform}),  the
slicing cuts are $\min(s_{q_1g_1},s_{g_1g_2}) < \delta$ and the
subtraction occurs over the range  $\delta <
\min(s_{q_1g_1},s_{g_1g_2}) < \Delta$.  Similar transformations and
cuts act over the other two antenna.

\subsubsection{Slicing contribution}

For the five parton matrix elements of eq.~(\ref{eq:sig5}), the sum of
infrared singularities from the three antennae in the slicing approach
gives a contribution to the four particle final state which can be read
directly from eq.~(3.79) of ref.~\cite{GG},
\begin{equation}
 \d \sigma_{4}^{\rm slice} 
= R(Q_1;G_1,G_2;\Qbar_2)  \d \sigma_{4}^{\rm LO} .
\label{eq:4slice}
\end{equation}

Retaining only the leading colour contribution (i.e. dropping the 
contributions from the four quark process proportional to the number of
quark flavours),  
\begin{eqnarray} 
R(Q_1;G_1,G_2;\Qbar_2)   & = &
\left(\frac{\alpha_sN}{2\pi}\right)\frac{1}{\Gamma (1-\e )}~ 
\Biggl
\lbrack  \sum_{ij} \left\{
\frac{1}{\e^2}\left(\frac{4\pi\mu^2}{s_{ij}}\right)^{\e} -\log^2
\left(\frac{s_{ij}}{\delta}\right) \right\}\nonumber\\ 
&&+\frac{3}{2\e}\left(\frac{4\pi\mu^2}{\delta}\right)^\e
+\frac{197}{18}-\pi^2 \Biggr \rbrack \nonumber \\
&+&\as\frac{2b_0}{\e}\frac{1}{\Gamma(1-\e)}\left(\frac{4\pi\mu^2}{\delta}\right)^\e
+{\cal O}(\e) + {\cal O}(\delta),\nonumber 
\label{eq:Rfac}\end{eqnarray} 
with (at leading order in the number of colours)
$b_0=11N/6$ and where  the sum runs over the pairs of adjacent (colour
connected) hard partons, $ij=Q_1G_1$, $G_1G_2$ and $G_2\Qbar_2$.

\subsubsection{Subtraction term}

Since there are three antennae, we subtract three antennae factors,
such that the total subtraction term is, 
\begin{eqnarray} 
\frac{\d\sigma_{5}^{\rm sub}}{\sigma_0} &=&\frac{(2\pi)^7}{s}
\left(\frac{N^2-1}{N^2}\right) \asN^{3} 
 ~\d PS (Q^2;q_1,g_1,g_2,g_3,\bar q_2)\nonumber \\ 
&\times& \Biggl (
~\A_{q_1g_1g_2}
\Big |\S_\mu(Q_1;G_1,g_3;\bar q_2) V^\mu\Big |^2~\J_{(4)} \nonumber \\ 
&& +~ \A_{g_1g_2g_3}
\Big |\S_\mu(q_1;G_1,G_2;\bar q_2) V^\mu\Big |^2~\J_{(4)} \nonumber \\ 
&& +~\A_{g_2g_2\bar q_2}
\Big |\S_\mu(q_1;g_1,G_2;\Qbar_2) V^\mu\Big |^2~\J_{(4)} ~ \Biggr ) . 
\label{eq:sub5}  
\end{eqnarray}  
Here, we have used the mappings  $\{p_{q_1},p_{g_1},p_{g_2}\} \to
\{p_{Q_1},p_{G_1}\}$ according to eq.~(\ref{eq:transform})  for the
first antenna. We recall that the subtraction occurs over the range
$\delta < \min(s_{q_1g_1},s_{g_1g_2}) < \Delta$ and that the observable
function $\J_4$  is applied to the momenta for $Q_1$, $G_1$, $g_3$ and
$\bar q_2$. Similar procedures are applied to the other antennae.

However, we must add these terms back to the four parton contribution.
Here it is simplest to re-identify each of the four particle momenta
with the momenta relevant for tree level. In other words, for the first
antenna, $\{p_{q_1},p_{g_1},p_{g_2}\} \to \{p_{Q_1},p_{G_1}\}$ as
before and $p_{g_3} \to p_{G_2}$, $p_{\bar q_2} \to p_{\Qbar_2}$. This
is safe to do since we integrate over the whole four particle phase
space. Altogether, we have,
\begin{equation}
 \d \sigma_{4}^{\rm sub} 
=\left(\F_{Q_1G_1}+\F_{G_1G_2}+\F_{G_2\Qbar_2}\right)
 \d \sigma_{4}^{\rm LO} .
\label{eq:sub4} 
\end{equation} 

\subsubsection{Virtual contribution}

From ref.~\cite{us}, the matrix elements for the leading colour
one-loop contribution to the $q\bar q gg$ final state for this colour
ordering can be written in terms of a pole structure in $\e$ 
multiplying the lowest order matrix elements and a function $\Lhat_A$
that is finite as $\e \to 0$, 
\begin{equation} 
\L_A(G_1,G_2) =  
V(Q_1;G_1,G_2;\Qbar_2) \Big |\S_\mu(Q_1;G_1,G_2;\Qbar_2) V^\mu\Big |^2 
+ \Lhat_A(G_1,G_2).  
 \end{equation} 
The divergent factor $V$ is given
by, 
\begin{equation} 
V(Q_1;G_1,G_2;\Qbar_2) = \asN \left(
-\frac{\P(s_{Q_1G_1})}{\e^2} - \frac{\P(s_{G_2\Qbar_2})}{\e^2} -
\frac{\P(s_{G_1G_2})}{\e^2} - \frac{3}{2} \frac{\P(Q^2)}{\e} \right)
\label{eq:Vfac} 
\end{equation} 
where we have introduced the notation,
\begin{equation} 
\P(s) = \left(\frac{4\pi\mu^2}{-s}\right)^{\e}
\frac{\Gamma^2(1-\e)\Gamma(1+\e)}{\Gamma(1-2\e)}. 
\end{equation} 
In
terms of cross sections, we have, 
\begin{equation} 
\d \sigma_{4}^{\rm
V}  = V(Q_1;G_1,G_2;\Qbar_2) \d \sigma_{4}^{\rm LO}  +  \d
\sigma_{4}^{\rm V,finite} , 
\end{equation} 
with, 
\begin{equation}
\frac{\d \sigma_{4}^{\rm V,finite}}{\sigma_0} = \frac{(2\pi)^5}{s}
\left(\frac{N^2-1}{N^2}\right) \asN^3 \Lhat_A(G_1,G_2) ~\J_{(4)}  ~\d
PS (Q^2;p_{Q_1},p_{G_1},p_{G_2},p_{\Qbar_2}) \label{eq:Vfinite}
\end{equation}

Here $\Lhat_A$ is a finite function given  in terms of that logarithms
and dilogarithms that arise in evaluating the one-loop
contributions.\footnote{Note that $\Lhat_A$ defined here is a factor of
8 smaller than that in \cite{us}.} We have ensured that the kinematic 
singularity structure of $\Lhat_A$ matches that  of the tree-level
$\Big |\S_\mu(Q_1;G_1,G_2;\Qbar_2) V^\mu\Big |^2$. It can be written
symbolically as, 
\begin{equation} 
\Lhat_A=\sum_i P_i(s) {\rm L}_i,
\end{equation} 
where the coefficients $P_i(s)$ are rational polynomials of 
invariants. The finite functions ${\rm L}_i$ are the linear
combinations of scalar integrals defined in \cite{us}  which are
well-behaved in all kinematic limits, so that $\Lhat_A$ is numerically
stable. Unfortunately, the analytic expression for $\Lhat_A$ is rather
lengthy so we do not reproduce it here.

\subsubsection{Next-to-leading order cross section}

Assembling the various pieces, and applying coupling constant
renormalisation, 
\begin{equation} 
\as \to \asmu
\left(1-b_0\asmu\frac{(4\pi)^{\e}}{\e\Gamma(1-\e}\right),
\end{equation} 
the NLO four parton contribution is, 
\begin{eqnarray} \d
\sigma_{4}^{\rm NLO}   &=& \d \sigma_{4}^{\rm V}  
+ \d \sigma_{4}^{\rm sub}  
+ \d \sigma_{4}^{\rm slice} \nonumber \\ 
&=&
K(Q_1;G_1,G_2;\Qbar_2)\d \sigma_{4}^{\rm LO} 
 +\d \sigma_{4}^{\rm V,finite}, 
\end{eqnarray} 
where $\d \sigma_{4}^{\rm LO}$ and $\d \sigma_{4}^{\rm V,finite}$  are
given by eqs.~(\ref{eq:LO}) and (\ref{eq:Vfinite}) respectively with
the replacement $\alpha_s \to \alpha_s(\mu)$. The factor $K$ is the sum
of the divergent one loop factor (eq.~(\ref{eq:Vfac})), the slicing
factor (eq.~(\ref{eq:Rfac})) and the subtraction term
(eq.~(\ref{eq:sub4})), 
\begin{eqnarray}
K(Q_1;G_1,G_2;\Qbar_2) &=& V(Q_1;G_1,G_2;\Qbar_2)
+R(Q_1;G_1,G_2;\Qbar_2)  \nonumber \\ 
&& +\F_{Q_1G_1}(s_{Q_1G_1})
+\F_{G_1G_2}(s_{G_1G_2}) +\F_{G_2\Qbar_2}(s_{G_2\Qbar_2})\nonumber \\
&=& \asNmu \Biggl( \frac{197}{18} + \frac{\pi^2}{2} \nonumber \\ 
&&~~~
+\F_{Q_1G_1}^\Delta\left(\frac{\Delta}{s_{Q_1G_1}}\right)
+\F_{G_1G_2}^\Delta\left(\frac{\Delta}{s_{G_1G_2}}\right)
+\F_{G_2\Qbar_2}^\Delta\left(\frac{\Delta}{s_{G_2\Qbar_2}}\right) 
\nonumber \\ 
&&~~~ -\frac{10}{6}\log\left(\frac{s_{Q_1G_1}}{Q^2}\right)
-\frac{11}{6}\log\left(\frac{s_{G_1G_2}}{Q^2}\right)
-\frac{10}{6}\log\left(\frac{s_{G_2\Qbar_2}}{Q^2}\right) \Biggr )
\nonumber \\ 
&&~~~+ \asmu 2 b_0 \log\left(\frac{\mu^2}{Q^2}\right).
\end{eqnarray}

Similarly, the five parton leading colour 
contribution to four jet-like observables
is obtained from eqs.~(\ref{eq:sig5}) and (\ref{eq:sub5}), 
\begin{equation}
\d \sigma_{5}^{\rm NLO} 
=
 \d \sigma_{5}-\d \sigma_{5}^{\rm sub},
\end{equation}
evaluated with the running $\alpha_s(\mu)$. By construction this is
finite as any one particle becomes unresolved. In the slicing regions,
$\d \sigma_{5}^{\rm NLO} = 0$, while the phase space regions over which
the subtraction terms are  applied are implicit in the definition of
the antenna functions.

Note that the four-dimensional limit of all cross sections may be taken
with impunity now that the singularities have cancelled. Furthermore,
there is no dependence in $K$ on the slicing parameter $\delta$ which
may also be taken as small as desired.   The subtraction parameter
$\Delta$ remains, and both $\d \sigma_{4}^{\rm NLO}$ and $\d
\sigma_{5}^{\rm NLO}$ individually depend on it.   However, the sum of
both contributions is independent of the choice of $\Delta$.   The
precise value of $\Delta$ can be made bearing in mind the numerical
stability and speed of the final computer code.   For small $\Delta$,
there may be sizeable cancellations between  the four and five parton
contributions, while for large $\Delta$ more CPU time is required to
evaluate the subtraction terms. For our numerical results, we have
taken,
\begin{equation}
\delta = 10^{-8}, \qquad \qquad \Delta = 10^{-3}.
\end{equation}

\end{document}